\documentclass[final]{jfm}
\usepackage{natbib}
\usepackage{epsfig}
\usepackage{graphicx}
\usepackage{dcolumn}
\usepackage{bm}
\usepackage{setspace}
\usepackage{amsmath}
\usepackage{array}
\usepackage{url}
\usepackage{color}
\usepackage{rotating}

\begin{document}

\title[Dynamics of sessile drops. Part 3. Theory of forced oscillations]{Dynamics of sessile drops. \\ Part 3. Theory of forced oscillations }
\author[J.B. Bostwick and P.H. Steen]{J.\ns B.\ns B\ls O\ls S\ls T\ls W\ls I\ls C\ls K$^{1}$\footnote{Email address for correspondence: jbostwi@clemson.edu} \ns \and P.\ns H.\ns S\ls T\ls E\ls E\ls N$^{2,3}$}
\affiliation{$^1$Department of Mechanical Engineering, Clemson University, Clemson, SC 29631, USA\\ $^{2}$Department of Theoretical \& Applied Mechanics, Cornell University, Ithaca, NY 14853, USA \\  $^3$School of Chemical and Biomolecular Engineering and Center for Applied Mathematics, Cornell University, Ithaca, NY 14853, USA
}

\date{\today}
\maketitle

\begin{abstract}
A partially-wetting sessile drop is driven by a sinusoidal pressure field that produces capillary waves on the liquid/gas interface. The analysis presented in Part 1 of this series \citep{bostwickS} is extended by computing response diagrams and phase shifts for the viscous droplet, whose three phase contact-line moves with contact-angle that is a smooth function of the contact line speed. Viscous dissipation is incorporated through the viscous potential flow approximation and the critical Ohnesorge number bounding regions beyond which a given mode becomes over-damped is computed. Davis dissipation originating from the contact-line speed condition leads to damped oscillations for drops with finite contact-line mobility, even for inviscid fluids. The critical mobility and associated driving frequency to generate the largest Davis dissipation is computed. Lastly, regions of modal coexistence where two modes can be simultaneously excited by a single forcing frequency are identified. Predictions compare favorably to related experiments on vibrated drops.
\end{abstract}


\section{Introduction}
Driven droplets play a critical role in a number of emerging technologies, such as 3D printing \citep{calvert2001inkjet} with application to rapid prototyping \citep{grimm2004}, self-cleansing surfaces for enhanced solar cell efficiency \citep{park11}, microfluidics \citep{stone04}, inkjets \citep{CastrejonPita2013,basaran2013nonstandard}, spray cooling for high heat flux applications \citep{kim2007spray}, and drop atomization for drug delivery (aerosol) methods \citep{donnelly2004experimental}, all of which involve the motion of liquids on scales where surface tension dominates. 

Forcing of sessile drops can induce shape change or drive fluid transport. Shape change occurs in experiments by driving droplets using electrowetting \citep{mampallil2013electrowetting}, surface acoustic waves \citep{baudoin2012}, air jets \citep{deepu2014multimodal}, mechanically vibrated substrates \citep{smith,chang13} or pressure excitations \citep{tilger2013phase}.  Bulk translational motion of driven droplets can be achieved provided contact angle hysteresis can be overcome to mobilize the three phase contact line \citep{noblin2004vibrated}. \cite{brunet2007vibration} have demonstrated that a mechanically-vibrated drop can be made to walk `uphill' against gravity in a ratchet-like motion \citep{noblin2009ratchetlike}. A description of the fluid mechanical droplet response to the applied driving force is crucial in understanding the aforementioned applications, as well as guiding future studies led by prediction.

In Part 1 of this series \citep{bostwickS}, we analyzed the linear stability of the sessile drop, parameterized by the static contact angle $\alpha$, with a contact line that was i) pinned, ii) mobile or iii) obeyed a constitutive law that relates the contact angle to the contact line speed. The natural frequencies $\omega$ obey an operator equation
\begin{equation} -\omega^2 M[y] + K[y]= 0, \label{SHO} \end{equation}
for the disturbance shape $y$ that takes the familiar form for a simple harmonic oscillator with $M$ representative of fluid inertia (mass) and $K$ the restoring force of capillarity (spring constant). Eigenmode solutions of (\ref{SHO}) are characterized by the wavenumber pair $[k,l]$ that follow the spherical harmonic classification scheme \citep{MacRobert}. The associated eigenfrequencies depend strongly upon the wetting properties of the solid substrate $\alpha$ and the mobility of the contact-line $\Lambda$. Our frequency predictions compare favorably to experiments \citep{sharp11,chang13} and finite element simulations \citep{basaran} in the appropriate limits.

In this part, we extend our analysis by introducing external forcing through the drop's bulk pressure $p=F_0 \mathrm{e}^{i \lambda t}$ with $\lambda$ the applied forcing frequency, as is the standard approach for Faraday oscillations \citep{Benjamin1954}. When bulk viscosity is included in the model, our forced-damped extension takes the form of a damped-driven oscillator
\begin{equation}-\lambda^2 M[y] + i \lambda \epsilon C[y] + K[y] = F_0. \label{forcedDDO}\end{equation}
The dissipation $C$ encompasses bulk dissipation from viscosity $\epsilon$ and \cite{davis} dissipation related to the dynamic effects associated with the contact-line speed law, as outlined in \citet[][Sec. 3.3]{bostwickARFM}. For reference, this condition is sometimes referred to as the Hocking condition in the literature \citep{hocking1}. In Part 2 \citep{chang15}, we solved (\ref{forcedDDO}) for drops with pinned contact lines ($\Lambda=\infty$) using viscous potential flow to evaluate the dissipation $C$ for a specific $\epsilon=0.0024$ to compare with experiment. The focus in this part is the contact-line mobility and a comprehensive exploration of the parameter space. We solve (\ref{forcedDDO}) reporting response diagrams and phase shifts, as they depend upon the viscosity $\epsilon$, contact angle $\alpha$ and contact line mobility $\Lambda$.

Lord \cite{rayleigh} showed that free drops exhibit a discrete spectrum, which has come to be referred to as the Rayleigh-Lamb (RL) spectrum \citep{lamb}. The RL spectrum has been verified experimentally for free drops \citep{trinh,wang} and is relevant in applications where the drop may not be completely free \citep{noblin2,chebel}. Typical extensions for free drops include, but are not limited to, the effects of i) viscosity \cite[]{reid,chandrasekhar,miller,prosperetti}, ii) large-amplitude perturbations \cite[]{tsamopoulos,lundgren} or iii) constrained geometries \cite[]{strani,bostwick,ramalingam,bostwick3}. Recent experiments by \cite{chang13} have shown the inadequacy of the RL spectrum for partially-wetting drops ($\alpha=75^\circ$) with pinned contact lines. The theory developed in Part 1, which accounts for the wetting properties of the solid substrate, compares favorably with these experiments.

Forced drops exhibit a finite bandwidth of forcing frequencies over which a particular mode may be excited, in contrast to the discrete (delta-function) response for unforced drops. In Part 2 of this series \citep{chang15}, observed frequency bands in experiments on mechanically-excited sessile water drops over a range of static contact angles were reported. Viscosity tends to decrease the droplet amplitude response and increase the bandwidth for a given mode, consistent with the experiments by \cite{sharp12}. 

Here we also include viscosity through the viscous potential flow approximation \citep{joseph2}, in which the bulk dissipation is evaluated using the velocity potential. Our bandwidth predictions compare favorably against experiment over a range of contact angles \citep{sharp12}. In addition, we compute the critical viscosity $\epsilon_c$ above which the oscillations for a particular mode become overdamped. This represents a bound above which a given mode cannot be harmonically excited. Our results may be viewed as a guide in designing experiments or applications where excitation of a specific mode is desirable.

Davis dissipation occurs for drops with finite contact-line mobility and leads to attenuated droplet response and increased bandwidth, even for inviscid fluids \citep{lyubimov2,lyubimov1}. However, the scaling of Davis dissipation differs from that for bulk viscous effects. For reference, the decay rate $\gamma$ from viscous dissipation for a free drop scales with the viscosity $\nu$ as $\gamma=\nu/R^2 (k-1)(2k+1)$. We compute the critical mobility and forcing frequency to generate the largest Davis dissipation in order to guide future experiments. With regard to comparison against the \cite{chang15} experiments, we are able to reproduce the observed frequency envelopes only by considering effects from finite contact-line mobility, which strongly argues for treating similar problems using the contact-line speed law.

In Part 1, it was shown that spectral ordering for the sessile drop can become broken and disordered for a range of contact angles. For the forced problem with finite bandwidth, we showed that two distinct modes may be simultaneously excited by a single forcing frequency and mapped these regions of modal coexistence in parameter space for a number of modal pairs. Identifying these regions is important for 3D printing \cite{calvert2001inkjet}, mixing \citep{mampallil11} and drop atomization \citep{tsai2012ejection} applications. The contact-line mobility strongly affects the size of the coexistence regions. In Part 2, we showed that mode selection in experiment is hysteretic in the coexistence regimes. That is, the dominant mode possibly depends upon the direction of the frequency sweep. Here we focus on generating `operating windows' for particular droplet behavior in anticipation of future studies.

We begin this paper by deriving the governing equations for the forced droplet and show how the fluid response depends upon the wetting properties through the contact angle $\alpha$, contact line mobility $\Lambda$ and viscosity through the Ohnesorge number $\epsilon$. Viscous dissipation is introduced via the viscous potential flow approximation leading to the damping of oscillations for $\epsilon<\epsilon_c$, where $\epsilon_c$ is the critical viscosity above which oscillations are overdamped for a particular mode. We then show finite contact line mobility $\Lambda$ leads to Davis dissipation and compute the critical mobility $\Lambda_m$ leading to the largest dissipation. Lastly, we show that two distinct modes may be simultaneously excited by a single forcing frequency and map these regions of modal coexistence in parameter space for a number of modal pairs. Comparison with relevant experiments is made when appropriate.

\section{Mathematical formulation}
\begin{figure}
\begin{center}
\includegraphics[width=0.54\textwidth]{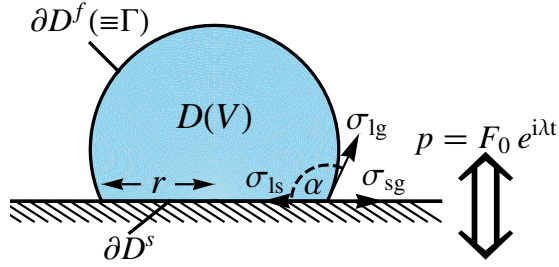}
\end{center}
\caption{\label{fig:SCdefsketch}Definition sketch: spherical-cap droplet of volume $V$ and free surface $\Gamma$ with contact angle $\alpha$ and liquid/gas $\sigma_{lg}$, solid/gas $\sigma_{sg}$ and liquid/solid $\sigma_{ls}$ surface tensions, driven by an applied pressure field $p=F_0 \mathrm{e}^{i\lambda t}$ of amplitude $F_0$ and frequency $\lambda$.  }
\end{figure}
Our derivation follows the boundary integral approach of \citet[][Sec. 1]{bostwick4} in which normal modes are invoked and the flow problem (interior domain) is mapped onto the undisturbed interface. For brevity, we follow the development set forth in Part 1 \citep{bostwickS}, extending the analysis presented in Part 2 \citep{chang15} to include the effects of contact-line mobility.

Consider an incompressible, viscous fluid subject to a time-dependent pressure field $p(t)=P_0\mathrm{e}^{i \Omega t}$, occupying a domain $D$ bounded by a spherical-cap interface $\partial D^f$ held by a constant surface tension $\sigma$ and a support surface $\partial D^s$, as shown in Figure~\ref{fig:SCdefsketch}. The equilibrium surface $\Gamma$ is defined parametrically as
\begin{equation} X(s,\varphi;\alpha)=\frac{\sin (s)}{\sin (\alpha)} \cos (\varphi),  \, Y(s,\varphi;\alpha)=\frac{\sin (s)}{\sin (\alpha)} \sin (\varphi),  \, Z(s;\alpha)= \frac{\cos (s) - \cos (\alpha)}{\sin (\alpha)} ,\label{SCeq}\end{equation}
using arclength-like $s\in \left[0,\alpha\right]$ and azimuthal angle $\varphi \in \left[0,2\pi\right]$ as surface coordinates, with $\alpha$ the static contact-angle. The interface is given a small perturbation $\eta(s,\varphi,t)$. No domain perturbation is needed for small deformations, thus the droplet domain
\begin{equation} D \equiv \{(x,y,z)|\,\,0 \leq x \leq X(s,\varphi;\alpha),\; 0 \leq y \leq Y(s,\varphi;\alpha),\;0\leq z \leq Z(s;\alpha)\} \label{SCdom}\end{equation}
is bounded by a free surface $\partial D^f \left(\equiv \Gamma\right)$ of constant surface tension $\sigma$, and a planar surface-of-support $\partial D^s$;
\begin{equation} \partial D^f\equiv\{\left(x,y,z\right)|\, x=X(s,\varphi;\alpha),\;y=Y(s,\varphi;\alpha),\; z=Z(s;\alpha)\}, \,\partial D^s\equiv\{\left(x,y,z\right)|\, z=0\}.  \label{SCsurf}\end{equation}

\subsection{Governing hydrodynamic equations}
We assume the velocity field $\boldsymbol{v}=-\boldsymbol{\nabla}\Psi$ can be expressed using the velocity potential $\Psi$ \cite[]{joseph}, noting that this form of the velocity field cannot satisfy the no-slip condition on the solid support, but we can evaluate the bulk dissipation from the irrotational field. This assumption is the essence of viscous potential flow theory. The velocity potential $\Psi$ satisfies the following boundary value problem,
\begin{equation} \nabla^2 \Psi = 0 \;\;\; [D],\quad\boldsymbol{\nabla}\Psi \cdot \hat{z} = 0 \;\;\; [\partial D^s],\quad\frac{\partial \Psi}{\partial n} = - \frac{\partial \eta}{\partial t} \;\;\; [\partial D^f]. \label{BVP}\end{equation}
The pressure field $p$ is given by the linearized Bernoulli equation
\begin{equation} p = \varrho \frac{\partial \Psi}{\partial t}+P_0 \mathrm{e}^{i\Omega t} \;\;\; [D],\label{SCbernoulli}\end{equation}
where $\varrho$ is the fluid density. Finally, disturbances to the equilibrium surface $\Gamma$ generate pressure gradients, and thereby flows, according to the Young-Laplace equation
\begin{equation} p-\mu\,\hat{n}\cdot \left(\nabla\otimes\nabla\Psi\right)\cdot\hat{n} =-\sigma\left(\Delta_{\Gamma}\eta+\left(\kappa_1^2+\kappa_2^2\right)\eta\right) \;\;[\partial D^f] , \label{SCYL1}\end{equation}
where $\otimes$ is the tensor product and $\mu$ the fluid viscosity. The Laplace-Beltrami operator $\Delta_{\Gamma}$ is defined on the equilibrium surface $\Gamma$ and operates on functions $\eta$,
\begin{equation} \Delta_{\Gamma}\eta \equiv \frac{1}{\sqrt{g}}\frac{\partial}{\partial u^{\mu}}\left(\sqrt{g}g^{\mu \nu} \frac{\partial \eta}{\partial u^{\nu}}\right) \label{surfLap}\end{equation}
with the surface metric given by
\begin{equation}g_{\mu \nu} \equiv \boldsymbol{x}_{\mu}\cdot \boldsymbol{x}_{\nu} = \left( \begin{array}{cc}
\mathrm{csc}^2(\alpha) & 0 \\
0 & \left(\mathrm{csc}(\alpha)\sin(s)\right)^2  \end{array} \right),\;\; g=\left(\sin(s)\mathrm{csc}^2(\alpha)\right)^2 ,\label{SCsurfmetric}\end{equation}
and $\mu, \nu = 1,2$, using notation standard to differential geometry \cite[e.g.][]{kreyszig}.

The governing equations (\ref{BVP})-(\ref{SCYL1}) are augmented with a boundary condition on the three-phase contact-line to yield a well-posed system of partial differential equations, a condition which we discuss later.

\subsection{Normal mode reduction}
We assume normal modes for the interface disturbance $\eta$ and velocity potential $\Psi$,
\begin{equation} \eta(s,\varphi,t) = y(s)\mathrm{e}^{i l \varphi}\mathrm{e}^{i \Omega t},\;\; \Psi(\boldsymbol{x},t)=\phi(\rho,\theta)\mathrm{e}^{i l \varphi}\mathrm{e}^{i \Omega t}, \label{SCNM}\end{equation}
with $l$ the azimuthal wavenumber and $\Omega$ the forcing frequency. The normal stress balance at the interface (\ref{SCYL1}) can be written as
\begin{equation} \sin^2(\alpha)\left(\left(\frac{\partial \phi}{\partial n}\right)''+\cot(s)\left(\frac{\partial \phi}{\partial n}\right)'+\left(2-\frac{l^2}{\sin^2(s)}\right)\left(\frac{\partial \phi}{\partial n}\right)\right) =\lambda^2 \phi - i \lambda \epsilon \hat{n}\cdot\left(\nabla \otimes\nabla \phi\right)\cdot \hat{n} + \lambda F_0,\label{goveqn}\end{equation}
where $\epsilon \equiv \mu/\sqrt{\varrho r \sigma}$ is the Ohnesorge number, $\lambda\equiv \Omega \sqrt{\varrho r^3/\sigma}$ the scaled forcing frequency, $F_0=P_0 r^2/\sigma$ the scaled forcing amplitude and $'=\mathrm{d}/\mathrm{d}s$. The contact-line dynamics obey the general contact-line law relating the deviation in contact-angle from its static value $\Delta\alpha$ to the contact-line speed $u_{CL}$ \citep[][\S 3.2.3, Fig. 1c]{bostwickS};
\begin{equation}\frac{\partial}{\partial s}\left(\frac{\partial \phi}{\partial n}\right)+\cos(\alpha) \left(\frac{\partial \phi}{\partial n}\right)=i\lambda \Lambda \left(\frac{\partial \phi}{\partial n}\right),\label{SCBCspeed}\end{equation}
where $\Lambda$ is the contact-line mobility \citep{davis,hocking1}. Note that $\Lambda=0$ corresponds to the natural and $\Lambda=\infty$ to the pinned contact-line disturbance, respectively. The velocity potential additionally satisfies the following auxiliary conditions \citep[][Eq. 2.13]{bostwickS};
\begin{equation} \nabla^2\phi-\frac{l^2}{\rho^2\sin^2\theta}\phi=0\;\;[D], \quad \frac{\partial \phi}{\partial n}=0 \;\; [\partial D^s], \quad \frac{\partial \phi}{\partial n}=-i\lambda y \;\; [\partial D^f], \quad\int_{\Gamma}{\frac{\partial \phi}{\partial n}\mathrm{d}\Gamma}= 0. \label{auxeq}\end{equation}

\subsection{Derivation of integrodifferential equation}
We write the solution to (\ref{goveqn}--\ref{auxeq}) as an integral equation
\begin{equation}\left(1-b^2\right) \frac{\partial \phi}{\partial n}(x) = - i\lambda \epsilon \int_b^1{G(x,y)\left(\hat{n}\cdot\left(\nabla \otimes\nabla \phi\right)\cdot \hat{n}\right)\mathrm{d}y} + \lambda^2\int_b^1{G(x,y) \phi(y) \mathrm{d}y} + F_0 \lambda \int_b^1{G(x,y)\mathrm{d}y},\label{goveqnsol} \end{equation}
using the Green's function
\begin{equation} G\left(x,y;l,\lambda,\Lambda\right)=\begin{cases} \xi(l)y_1(y;l)\left[\frac{\tau_2}{\tau_1}y_1(x;l)-y_2(x;l)\right] & b < x < y < 1 \\ \xi(l)y_1(x;l)\left[\frac{\tau_2}{\tau_1}y_1(y;l)-y_2(y;l)\right] &  b < y < x < 1 ,
 \end{cases} \label{SCgreenpin}\end{equation}
where  $x\equiv \cos(s) ,\, b\equiv \cos(\alpha)$.
The functions $y_1$ and $y_2$ belong to the kernel of the curvature operator $K$ and are given by
\begin{equation} \begin{split} y_1(x;0)=P_1(x), \, y_2(x;0)=Q_1(x) ,\, y_1(x;1)=P^{(1)}_1(x),\, y_2(x;1)=Q^{(1)}_1(x) , \\ y_1(x;l\geq2)=\left(x+l\right)\left(\frac{1-x}{1+x}\right)^{l/2},\, y_2(x;l \geq 2)=\frac{\left(x+l\right)}{2l\left(l^2-1\right)}\left(\frac{1+x}{1-x}\right)^{l/2} , \end{split} \label{SCcurvsol}\end{equation}
where $P_1,Q_1$ and $P^{(1)}_1,Q^{(1)}_1$ are the order $0$ and $1$ Legendre functions of index $1$, respectively \cite[]{MacRobert}. Similarly, the scale factor is given by
\begin{equation} \xi(l)\equiv\begin{cases} 1/2& l=1 \\ 1 &  l\neq 1,
 \end{cases} \label{SCwron}\end{equation}
while
\begin{equation}\tau_1=y'_1(b;l)+\left(\frac{b}{\sqrt{1-b^2}}-i \lambda \Lambda\right)y_1(b;l),\;\; \tau_2=y'_2(b;l)+\left(\frac{b}{\sqrt{1-b^2}}-i \lambda \Lambda\right)y_2(b;l). \label{SCtauang}\end{equation}
Note that the Green's function is parameterized by azimuthal wavenumber $l$, forcing frequency $\lambda$ and contact-line mobility $\Lambda$.

\subsubsection{Spectral reduction}
A solution series
\begin{equation} \phi = \sum_{j=1}^N {a_j \phi_j},\label{SCseries}\end{equation}
is applied to (\ref{goveqnsol}) and inner products are taken to generate a set of algebraic equations
\begin{equation}\sum_{j=1}^N {\left(m_{ij}+i\epsilon \lambda \tau_{ij}-\lambda^2\kappa_{ij}\right)a_j}=F_0 \lambda \gamma_i, \label{matrixeqn}\end{equation}
with
\begin{equation}\begin{split}m_{ij}\equiv \left(1-b^2\right)\int_{b}^{1}{\frac{\partial \phi_i}{\partial n}\phi_j\mathrm{d}x}, \quad \tau_{ij}\equiv \int_{b}^{1}\int_{b}^{1}{G(x,t)\left(\hat{n}\cdot\left(\nabla \otimes\nabla \phi_i\right)\cdot \hat{n}\right)\phi_j(x)\mathrm{d}x\mathrm{d}t}, \\ \kappa_{ij}\equiv \int_{b}^{1}\int_{b}^{1}{G(x,t)\phi_i(t)\phi_j(x)\mathrm{d}x\mathrm{d}t},  \quad \gamma_i\equiv \int_{b}^{1}\int_{b}^{1}{G(x,t)\phi_i(x)\mathrm{d}x\mathrm{d}t}.\end{split} \end{equation}
The auxiliary conditions (\ref{auxeq}) are satisfied through proper selection of the basis functions $\phi_j$, as discussed in \citet[][\S 4.2]{bostwickS}. For zonal modes,
\begin{equation}\phi_j(\rho,\theta)=\rho^{2j}P_{2j}\left(\cos\theta\right), \label{SCcoordfunc}\end{equation}
while for non-zonal modes,
\begin{equation}\phi^{(l)}_j(\rho,\theta)=\rho^{j}P^{(l)}_{j}\left(\cos\theta\right) \label{SCcoordfuncl}\end{equation}
with $j+l=\textrm{even}$.


\section{Results}

\begin{figure}
\begin{center}
\includegraphics[width=0.83\textwidth]{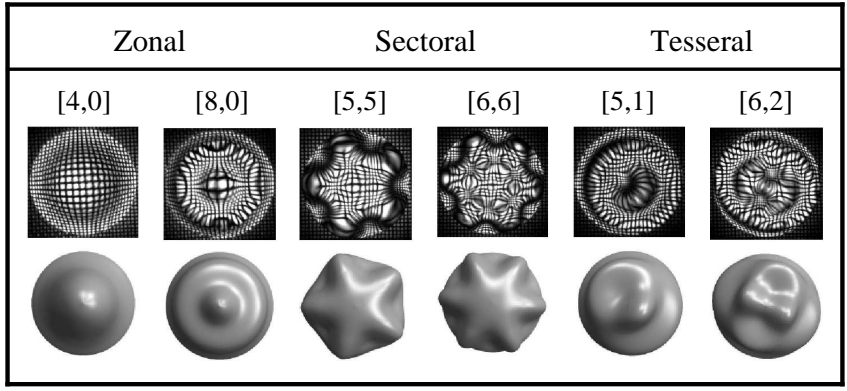}
\end{center}
\caption{\label{fig:modeID}Modal classification by wavenumber pair $[k,l]$ into zonal $[k,0]$, sectoral $[k,k]$ and tesseral $[k,l\neq k]$ shapes. Experimental images are reproduced from Part 2 \citep{chang15}.  }
\end{figure}

For fixed $\lambda,\epsilon,\alpha,l$, we compute the solution $a_j$ to the matrix equation (\ref{matrixeqn}). The associated fluid response $\phi, \partial \phi/\partial n$ is then obtained by applying $a_j$ to (\ref{SCseries}). Modal identities are distinguished by the wavenumber pair $[k,l]$ that follow the spherical harmonic classification scheme; zonal $[k,0]$, sectoral $[k,k]$ and tesseral $[k,l\neq k]$ shapes, as shown in Figure~\ref{fig:modeID}. An alternate identification uses layers and sectors \citep{chang15}. The focus here is the droplet response $a_j$, which is linear in the applied pressure amplitude $F_0$. Henceforth, we report the complex response as $c_j\equiv a_j/F_0$, which admits a phase shift
\begin{equation} \delta = \arctan{\Big|\frac{\mathrm{Im}[c]}{\mathrm{Re}[c]}\Big|}. \end{equation}
For $\delta=0^\circ$ and $\delta=90^\circ$, the droplet response is in-phase and out-of-phase with the applied pressure oscillations, respectively, with $\delta=90^\circ$ corresponding to a state of maximal dissipation. Note the damped-driven oscillator structure of (\ref{matrixeqn}) with corresponding features. In what follows, we show how the response diagram changes with viscosity $\epsilon$ and contact-line mobility $\Lambda$ and compare against relevant experiments when appropriate.

\subsection{Viscosity $\epsilon$}

\begin{figure}
\begin{center}
\begin{tabular}{cc}
Response & Phase shift  \\
\includegraphics[width=0.5\linewidth]{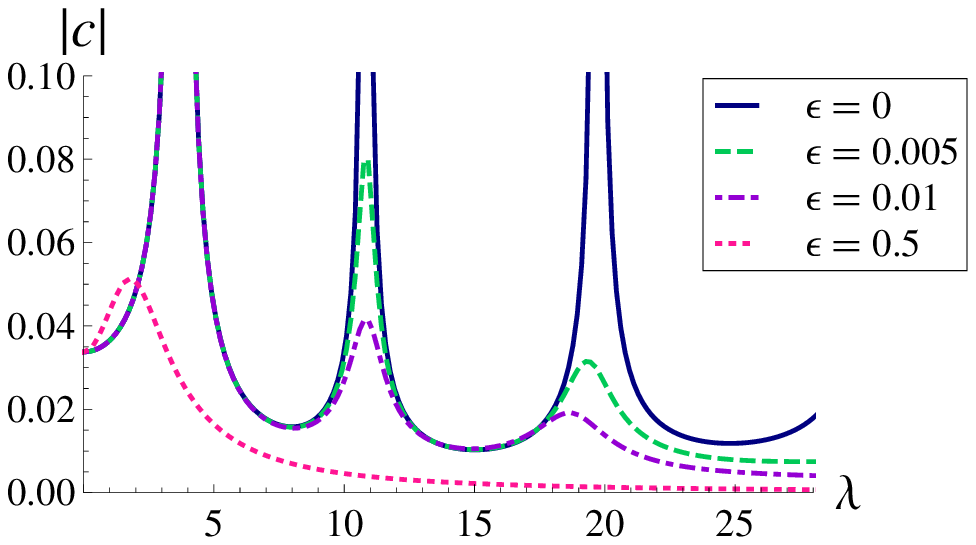} &
\includegraphics[width=0.5\linewidth]{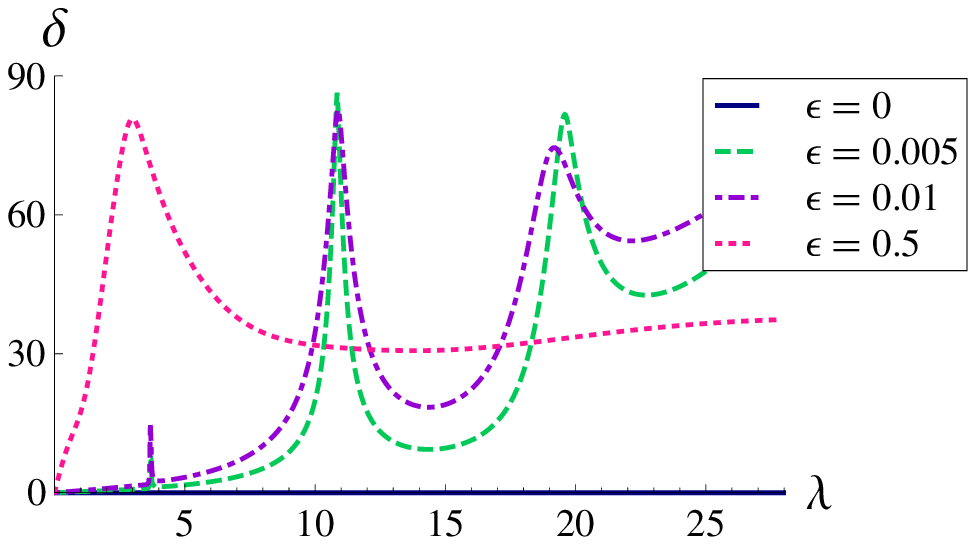}
\end{tabular}
\end{center}
\caption{Response $|c|$ (left) and phase shift $\delta$ (right) against frequency $\lambda$ for a drop with $\alpha=75^\circ$, $\Lambda=0$ (fully-mobile disturbance) and $l=0$, as it depends upon the Ohnesorge number $\epsilon$. The first peak corresponds to the $[2,0]$ mode, the second to $[4,0]$ and the third to $[6,0]$. Note for inviscid fluids $\epsilon=0$, the phase shift $\delta=0^\circ$ for all $\alpha$. \label{fig:viscresp} }
\end{figure}
Figure~\ref{fig:viscresp} plots the droplet response and phase shift for the zonal modes for a sub-hemispherical drop ($\alpha=75^\circ$) with $\Lambda=0$, as they depend upon the bulk viscosity $\epsilon$. For an inviscid fluid $\epsilon=0$, the oscillations are completely in phase $\delta=0^\circ$ with the applied field and the response diagram exhibits three infinite peaks that correspond to the $[2,0]$, $[4,0]$ and $[6,0]$ modes, respectively. Note that modes appear over a range of frequencies that define a bandwidth, a prominent feature of the forced oscillation problem that is also observed in experiment \citep{chang15}. For small viscosity $\epsilon=0.01$, the resonance peaks are dramatically lowered and the droplet response is out of phase $\delta\neq0^\circ$ with the driving frequency. The relative decrease in response amplitude $|c|$ for increasing wavenumber is consistent with bulk viscous effects leading to traditional viscous dissipation \citep{lamb}.

\begin{figure}
\begin{center}
\begin{tabular}{ccc}
\hspace{-0.3in}$\underline{l=0}$ &\hspace{-0.5in} $\underline{l=1}$ &\hspace{-0.5in} $\underline{l=2}$  \\
\includegraphics[width=0.37\linewidth]{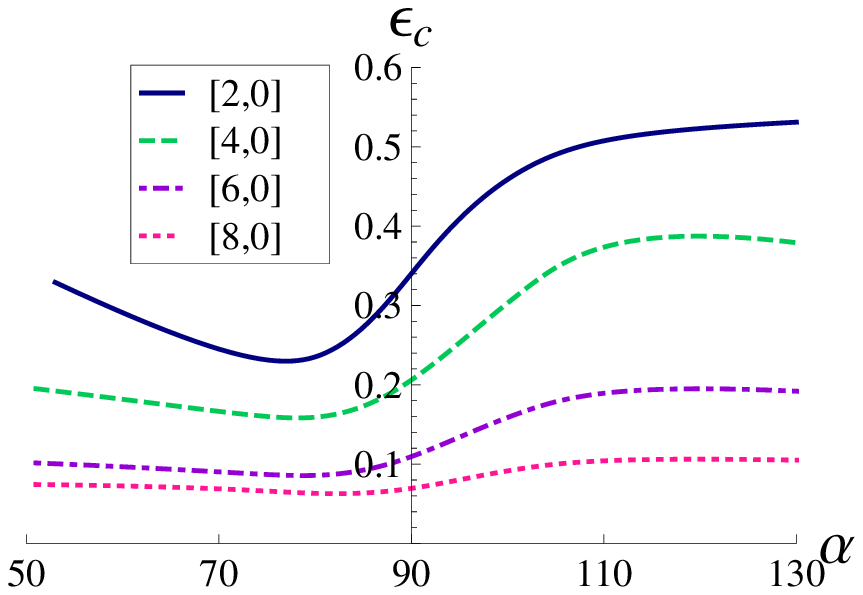}&
\hspace{-0.2in}\includegraphics[width=0.37\linewidth]{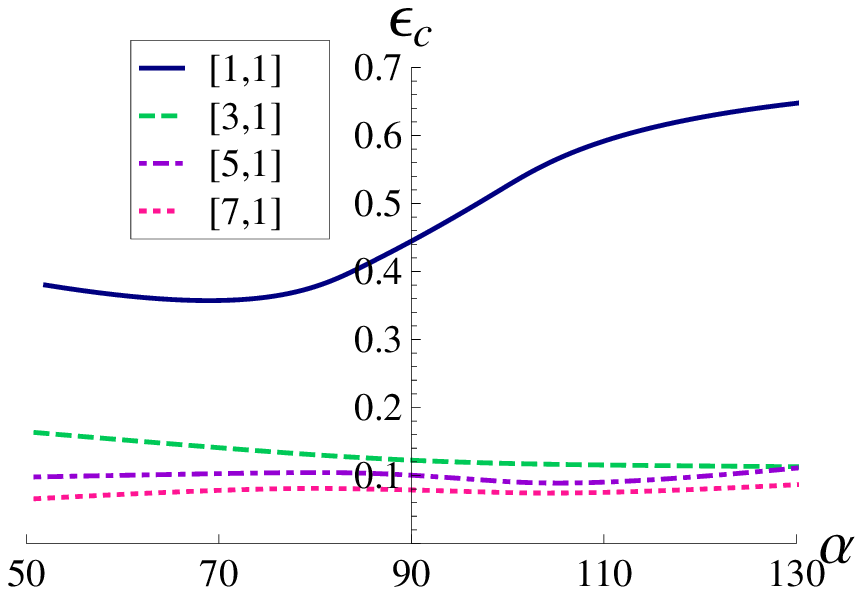} &
\hspace{-0.2in}\includegraphics[width=0.37\linewidth]{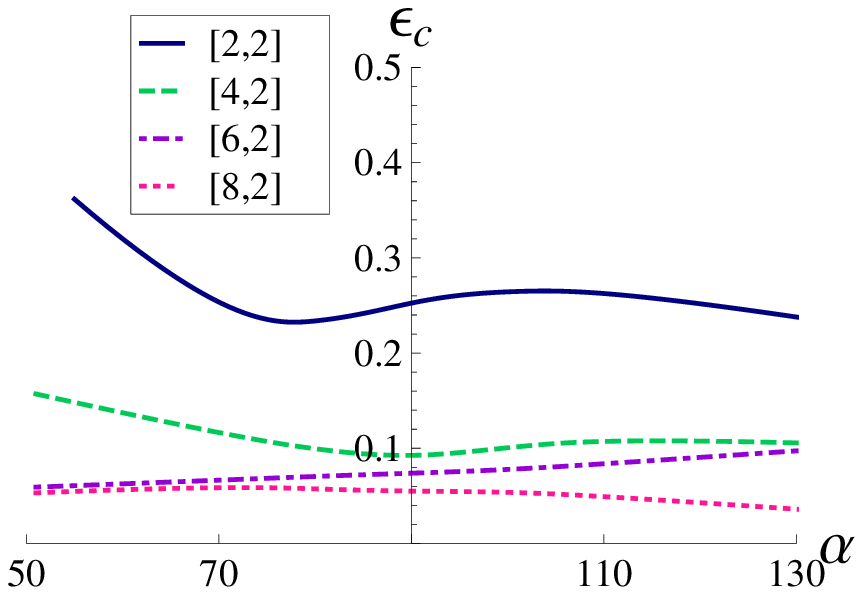} \\
\hspace{-0.3in}$\underline{l=3}$ & \hspace{-0.5in}$\underline{l=4}$ & \hspace{-0.5in}$\underline{l=5}$  \\
\includegraphics[width=0.37\linewidth]{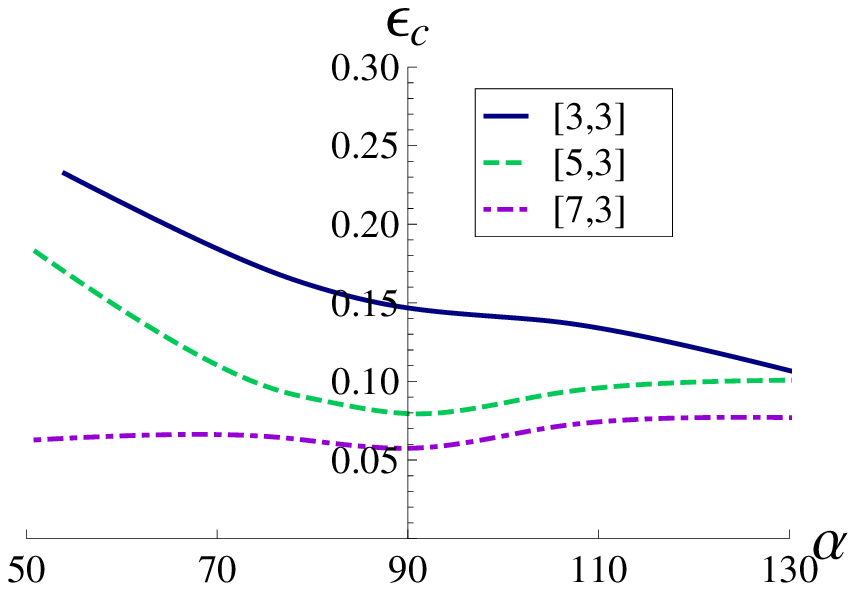}&
\hspace{-0.2in}\includegraphics[width=0.37\linewidth]{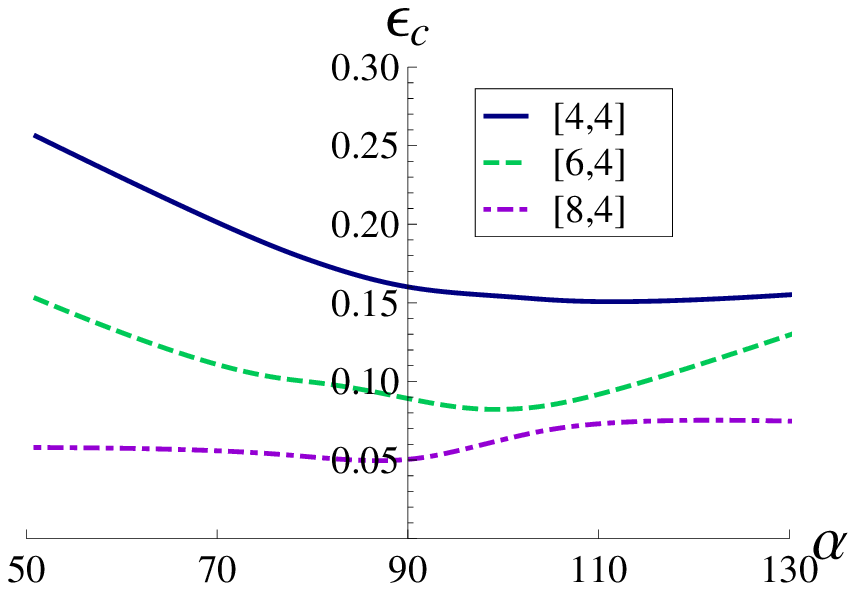} &
\hspace{-0.2in}\includegraphics[width=0.37\linewidth]{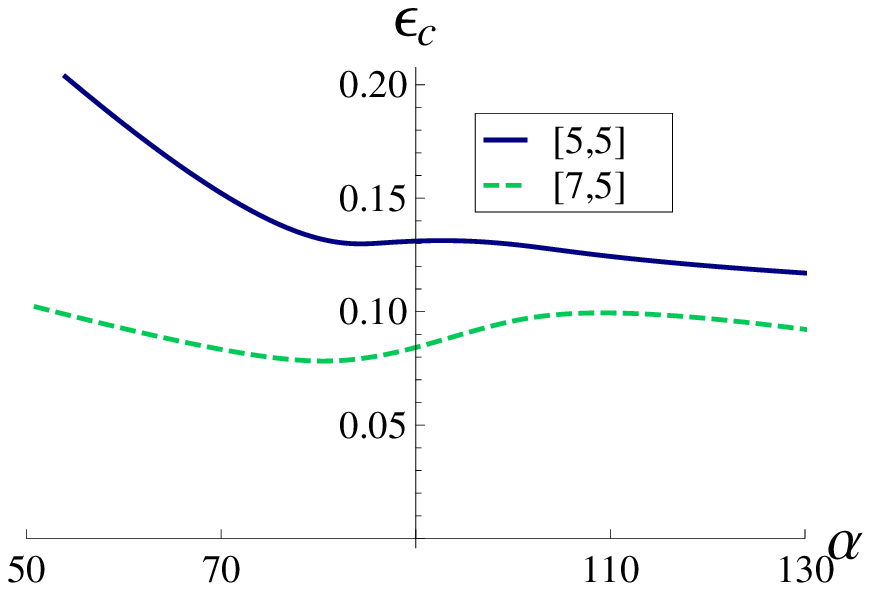}
\end{tabular}
\end{center}
\caption{Critical Ohnesorge number $\epsilon_c$ against contact-angle $\alpha$ for pinned disturbances $\Lambda=\infty$. Mode $[k,l]$ can be excited below the corresponding curve. Note the different vertical scales between sub-figures. \label{fig:critvisc} }
\end{figure}

Resonance peaks may disappear completely for large values of viscosity, as shown in Figure~\ref{fig:viscresp} for the $[4,0]$ and $[6,0]$ modes with $\epsilon=0.5$. For a given mode $[k,l]$, one can define a critical Ohnesorge number $\epsilon_c$ where the resonance peak disappears and above which ($\epsilon > \epsilon_c$) it is not possible to excite that mode. Stated differently, beyond $\epsilon_c$ the oscillations are over-damped. Figure~\ref{fig:critvisc} plots $\epsilon_c$ against contact angle for the pinned $\Lambda=\infty$ modes. Note that for a fixed azimuthal wavenumber $l$, $\epsilon_c$ decreases with increasing polar wavenumber $k$ irrespective of contact-angle, as could be expected from the increased surface distortion for the high wavenumber modes \citep[Fig. 7]{chang15}. However, the non-monotonic behavior with contact angle $\alpha$ could not have been predicted a priori and presumably results from the interactions between adjacent modes and the applied pressure field.

\begin{figure}
\begin{center}
\begin{tabular}{cc}
($a$)  & ($b$)  \\
\includegraphics[width=0.48\linewidth]{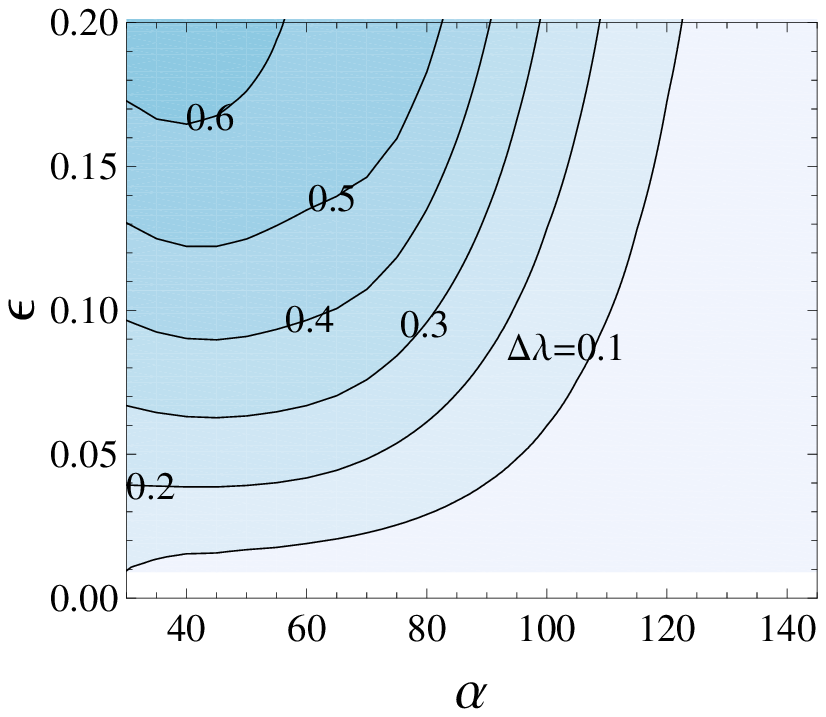} &
\includegraphics[width=0.56\linewidth]{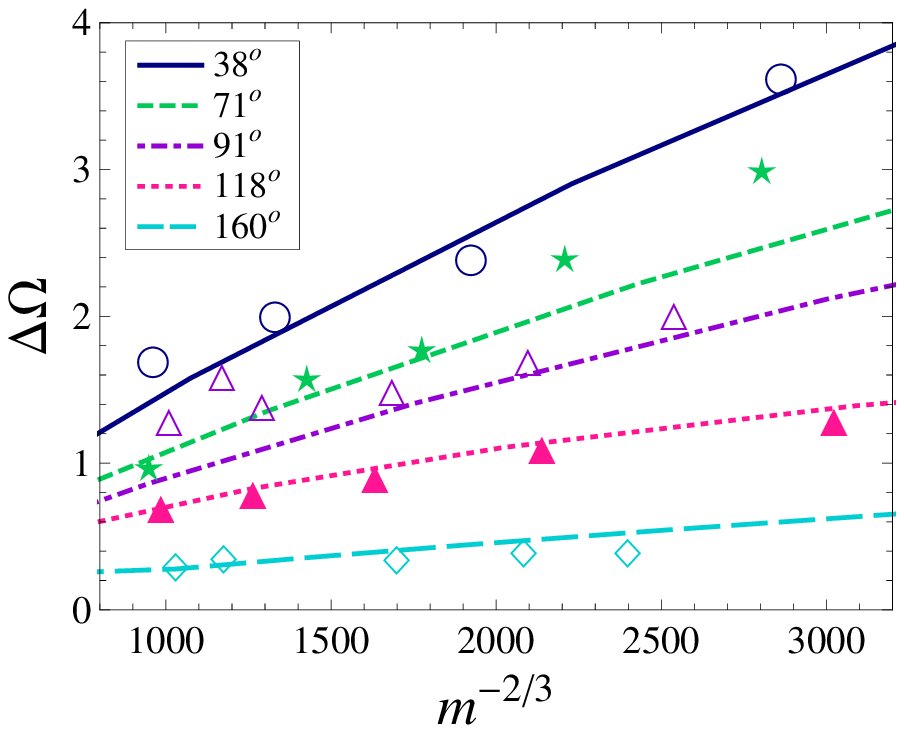}
\end{tabular}
\end{center}
\caption{Full width at half max (FWHM) for the $[1,1]$ pinned mode: ($a$) dimensionless frequency $\Delta\lambda$ against Ohnesorge number $\epsilon$ and contact-angle $\alpha$ and ($b$) dimensional frequency $\Delta\Omega$ against $1/m^{2/3}$ for 10\%w glycerol droplets comparing to \cite{sharp12} (symbols). \label{fig:FWHM}}
\end{figure}

A typical measure of the damping of oscillations in forced systems is the bandwidth of a resonance peak, which can easily be extracted from the response diagram (e.g.~Figure~\ref{fig:viscresp}). In particular, the full width at half max (FWHM) bandwidth also coincides with the decay rate of oscillations \citep{sharp12}. Figure~\ref{fig:FWHM}($a$) plots the dimensionless FWHM $\Delta\lambda$ against $\epsilon$ and $\alpha$ for the $[1,1]$ pinned mode. Note the non-monotonic dependence of the dissipation (FWHM) with respect to contact angle, reflecting the increased presence of the solid substrate for these wetting conditions. We compare our FWHM bandwidth predictions $\Delta\Omega$ for the $[1,1]$ pinned mode to the experiments by \cite{sharp12} over a wide range of contact angles in Figure~\ref{fig:FWHM}($b$). The agreement is reasonable over a large range of drop volumes, as measured by the drop mass $m$, suggesting the relevance of our theory to the given experiments.

\subsection{CL mobility $\Lambda$}

\begin{figure}
\begin{center}
\begin{tabular}{cc}
Response & Phase shift  \\
\includegraphics[width=0.5\linewidth]{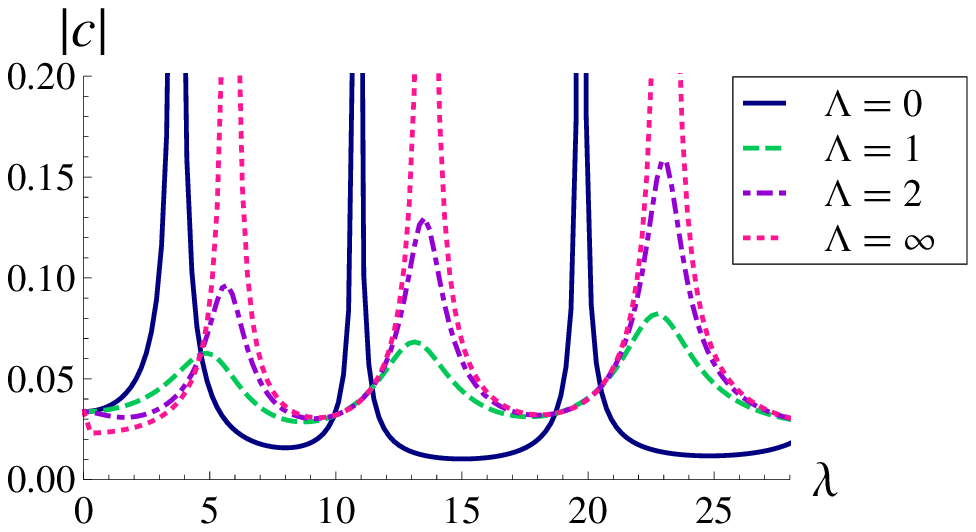} &
\includegraphics[width=0.5\linewidth]{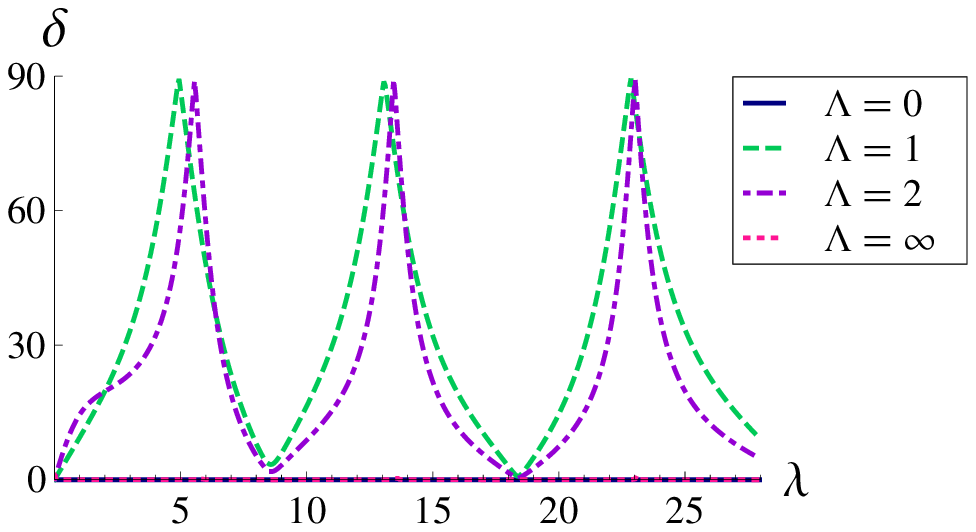}
\end{tabular}
\end{center}
\caption{Response $|c|$ (left) and phase shift $\delta$ (right) against frequency $\lambda$ for the zonal $l=0$ modes of an inviscid $\epsilon=0$ drop with $\alpha=75^\circ$, as it depends upon the mobility $\Lambda$. The first peak corresponds to the $[2,0]$ mode, the second to $[4,0]$ and the third to $[6,0]$. Note that the response is strictly real with phase shift $\delta=0^\circ$ for the free $\Lambda=0$ and pinned $\Lambda=\infty$ disturbances (superposed lines on $\lambda$ axis).  For finite mobility $\Lambda$, the drop response is out-of-phase $\delta=90^\circ$ at the resonant peak. \label{fig:mobresp}  }
\end{figure}

We examine the role of contact-line mobility by plotting the response diagram and phase shift for the zonal $l=0$ modes for an inviscid $\epsilon=0$ drop with $\alpha=75^\circ$ in Figure~\ref{fig:mobresp} for various values of $\Lambda$. We set $\epsilon=0$ to eliminate the effects of bulk viscous dissipation. For fully-mobile $\Lambda=0$ and pinned $\Lambda=\infty$ disturbances, the respective resonance peaks are infinite and the droplet oscillates in phase $\delta=0^\circ$ with the forcing frequency. However, for finite values of $\Lambda$ the resonance peak becomes finite and the oscillations become out-of-phase with the driving field, indicating that finite contact-line mobility $\Lambda$ leads to an effective dissipation. \cite{bostwickARFM} have deemed this feature `Davis dissipation' since it can be traced back to the work of \cite{davis} on fluid rivulets. Note that the resonance peaks become larger for increasing polar wavenumber $k$, indicating that the low order modes dissipate the most energy for fixed $\Lambda$. This feature is consistent with the results of \citet[][Fig. 13]{bostwickS}, who showed that for fixed azimuthal wavenumber $l$ the lower wavenumber $k$ modes have the largest contact-line excursion leading to increased Davis dissipation.

\begin{figure}
\begin{center}
\begin{tabular}{ccccc}
&\hspace{-0.35in} $\underline{k=1}$ & \hspace{-0.2in} $\underline{k=2}$ &\hspace{-0.2in} $\underline{k=3}$ & \hspace{-0.2in} $\underline{k=4}$ \\
\begin{sideways}\hspace{.35in} \large{$\Lambda_m$} \end{sideways} \hspace{0.4in}&
\hspace{-0.15in}\includegraphics[width=0.25\linewidth]{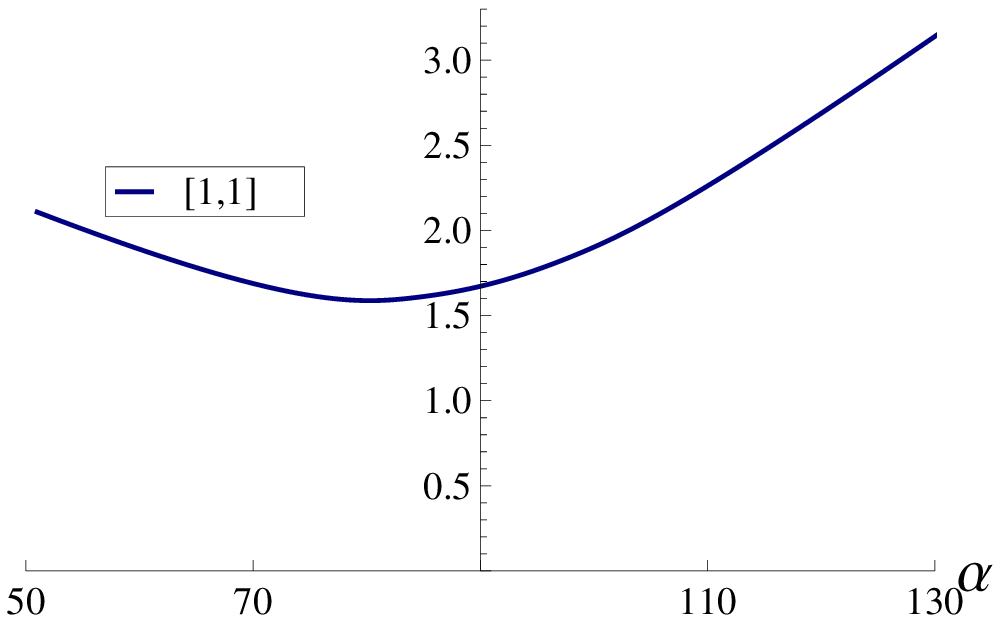}&
\hspace{-0.02in}\includegraphics[width=0.25\linewidth]{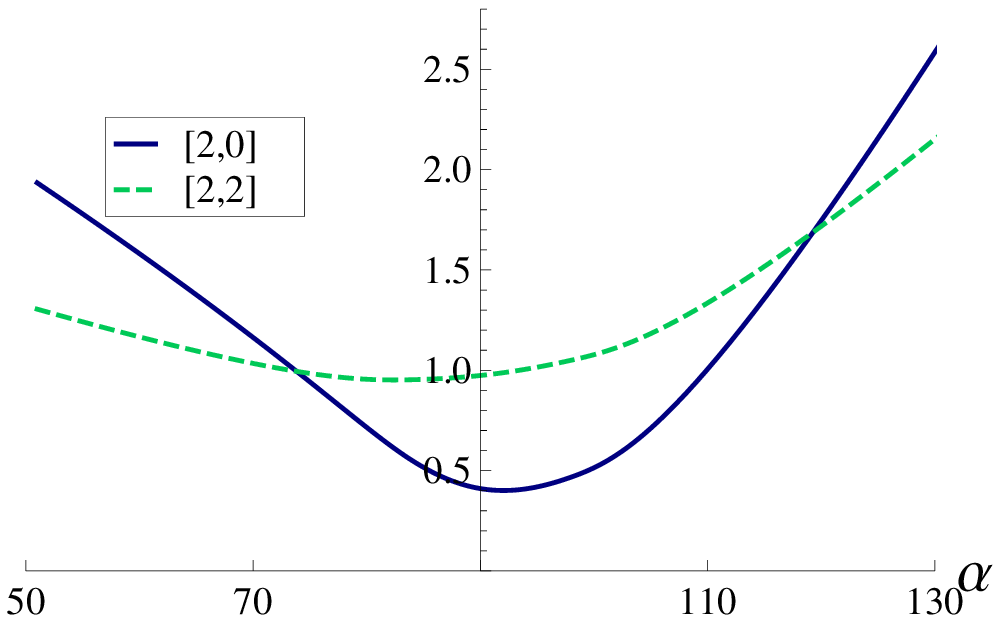} &
\hspace{-0.02in}\includegraphics[width=0.25\linewidth]{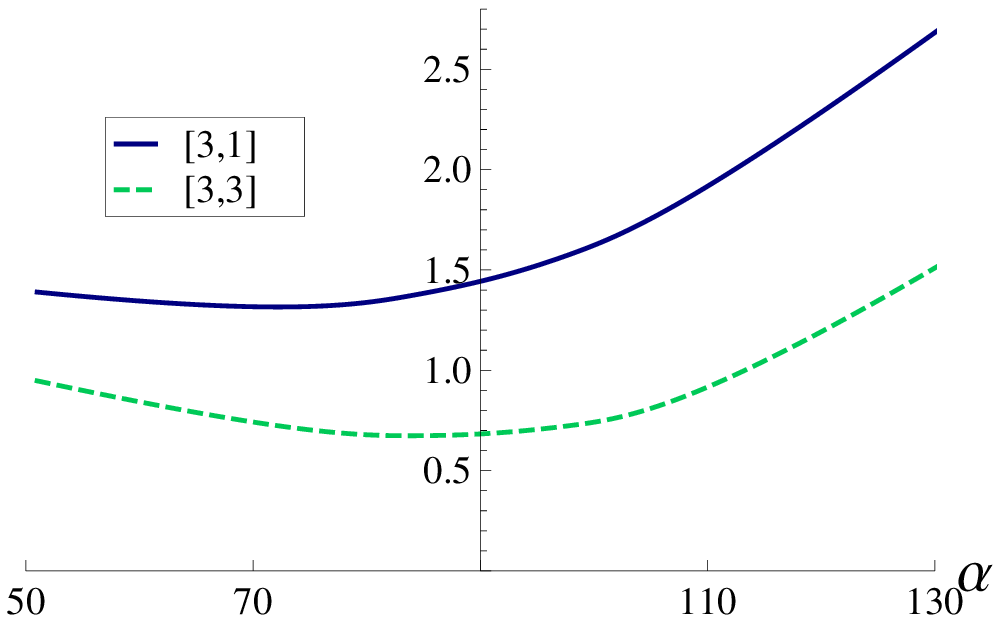} &
\hspace{-0.02in}\includegraphics[width=0.25\linewidth]{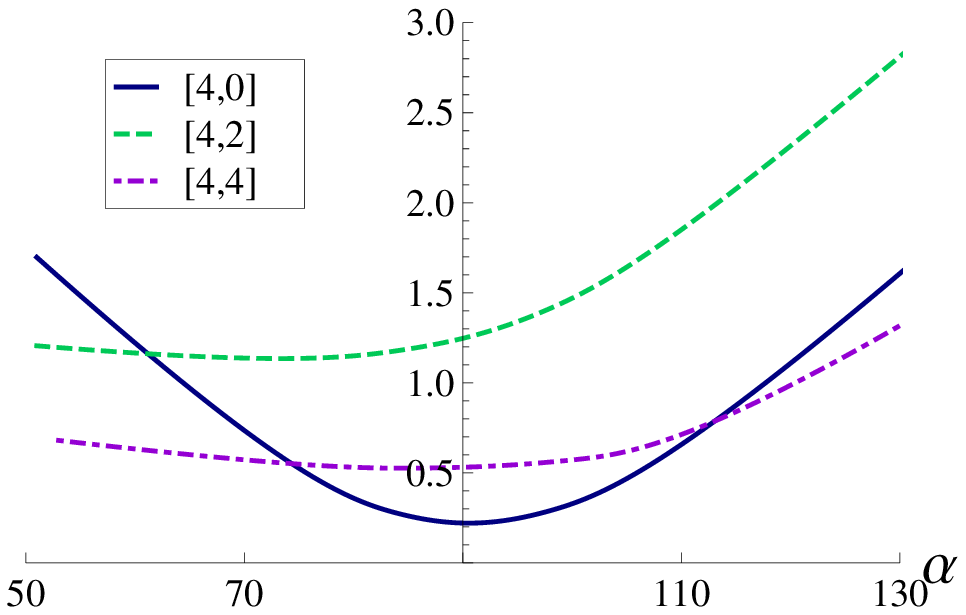}\\
\begin{sideways}\hspace{.35in} \large{$\lambda_m$} \end{sideways} \hspace{-0.15in}&
\hspace{-0.15in}\includegraphics[width=0.25\linewidth]{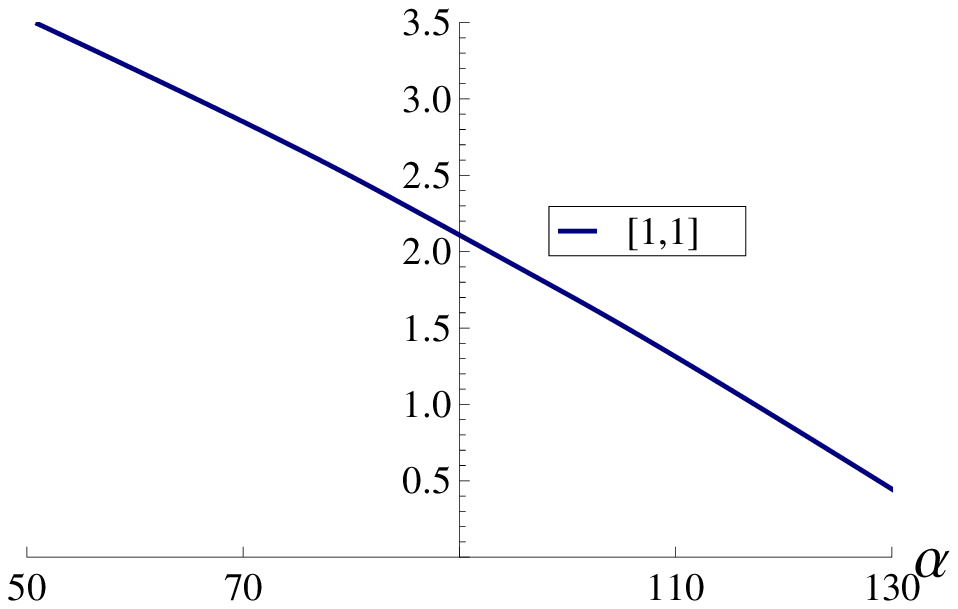}&
\hspace{-0.02in}\includegraphics[width=0.25\linewidth]{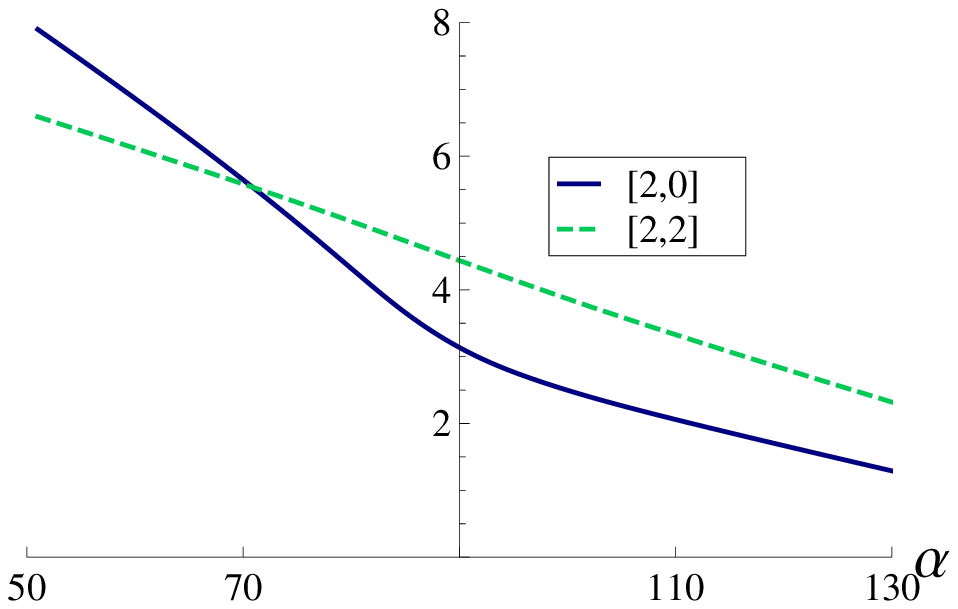} &
\hspace{-0.02in}\includegraphics[width=0.25\linewidth]{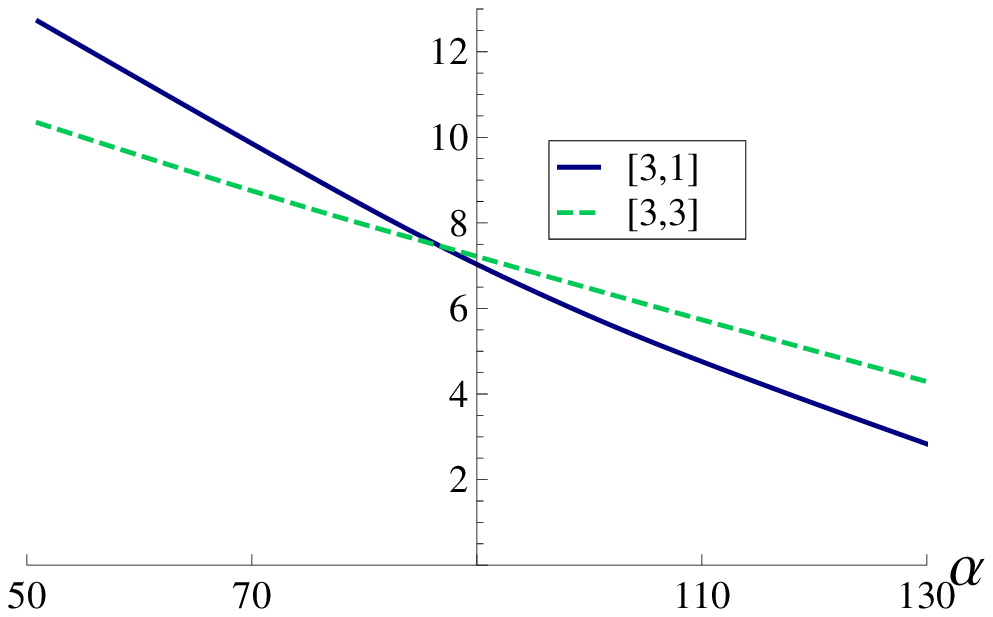} &
\hspace{-0.02in}\includegraphics[width=0.25\linewidth]{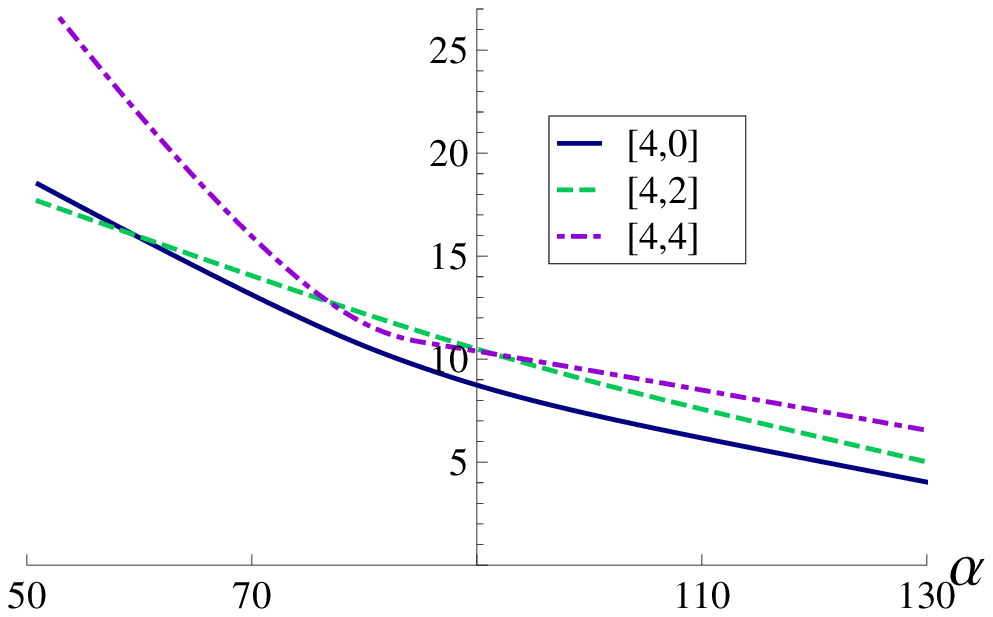}
\end{tabular}
\end{center}
\caption{Critical mobility $\Lambda_m$ and critical frequency $\lambda_m$ that generates the largest Davis dissipation for fixed polar wavenumber $k$, as it depends upon the contact-angle $\alpha$ and azimuthal wavenumber $l$. Viscous effects are negligible $\epsilon=0$. Note the different vertical scales between sub-figures. \label{fig:critmob}}
\end{figure}


For the mobility $\Lambda_m$, the resonance peak will be smallest and the droplet response is minimal. We call $\Lambda_m$ the critical mobility and $\lambda_m$ the critical frequency that generates the largest Davis dissipation. Figure~\ref{fig:critmob} plots $\Lambda_m, \lambda_m$ against contact-angle $\alpha$ for the $k=1-4$ modes. The forcing frequency $\lambda_m$ monotonically decreases with increasing contact angle, while the mobility $\Lambda_m$ is more complex. For example, the zonal modes $[2,0], [4,0]$ can have the smallest or largest critical mobility, for fixed polar wavenumber $k$, depending upon the contact angle. An important aspect of this study is the damping of oscillations for inviscid ($\epsilon=0$) fluids. Figure~\ref{fig:critmob} may be interpreted as a guide in selecting substrates for experiments that generate the largest Davis dissipation.

\begin{figure}
\begin{center}
\begin{tabular}{ccc}
$\underline{[5,5]}$ & $\underline{[7,7]}$ & $\underline{[9,9]}$  \\
\includegraphics[width=0.33\linewidth]{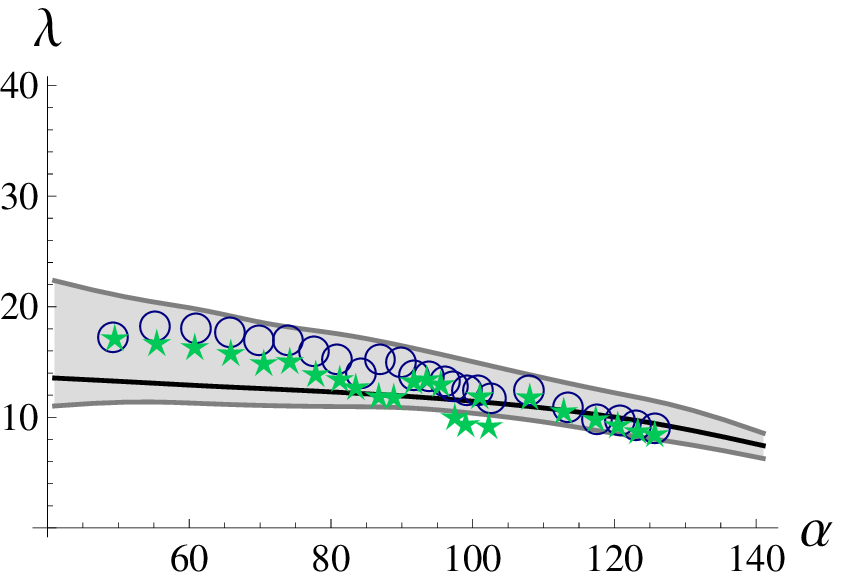}&
\includegraphics[width=0.33\linewidth]{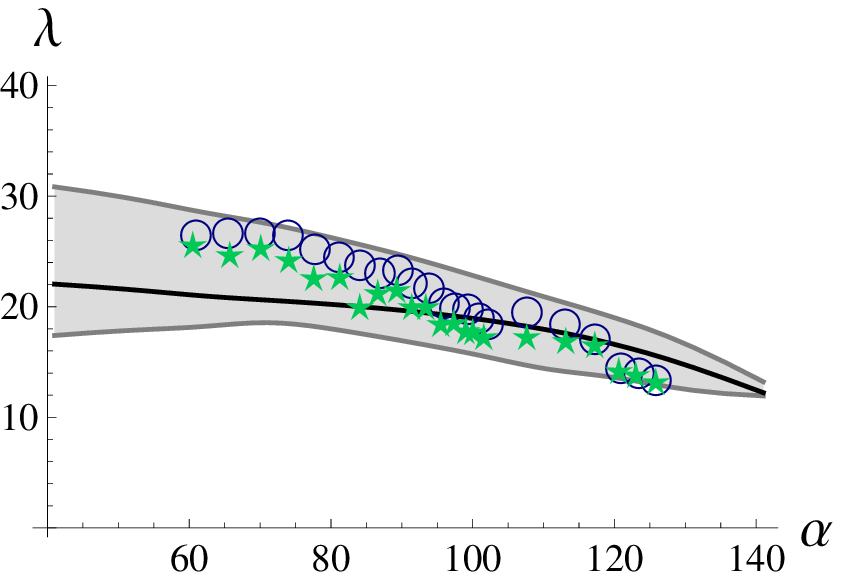} &
\includegraphics[width=0.33\linewidth]{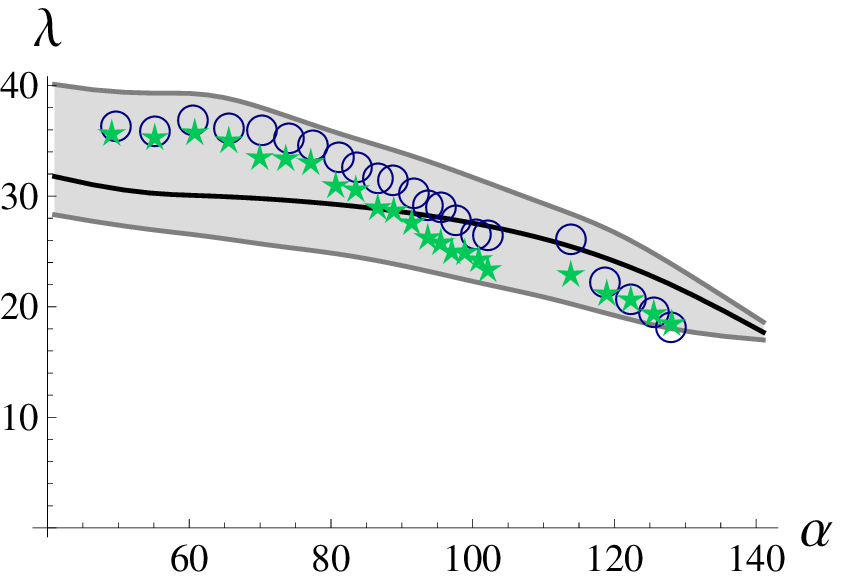}
\end{tabular}
\end{center}
\caption{Comparison with \cite{chang15} experiments: frequency envelopes against contact-angle $\alpha$ for sectoral modes $[5,5],[7,7],[9,9]$ with contact-line mobility $\Lambda=0.1$ and $\epsilon=0.0024$. Experiments given by symbols. \label{fig:changcomp}}
\end{figure}

The response diagrams of Figures~\ref{fig:viscresp},\ref{fig:mobresp}, show that modes can be excited over a range of forcing frequencies. This has been observed in recent experiments for a number of modes and over a range of contact angles \citep{chang15}. In that study, predicted frequency envelopes for pinned ($\Lambda=\infty$) disturbances with $\epsilon=0.0024$ compared favorably to experiment for a large number of modes, with the exception of the $[5,5], [7,7], [9,9]$ sectoral modes. By taking account of the contact-line mobility $\Lambda=0.1$, we are able to predict frequency envelopes that match the experiments for those remaining modes, suggesting that the contact-line dynamics are crucial in understanding the forced oscillations problem (c.f. Figure~\ref{fig:changcomp}).

\subsection{Modal coexistence}

An important prediction from Part 1 was that two modes with different wavenumber pair $[k,l]$ may share the same natural frequency and that the classical ordering of frequencies by increasing polar wavenumber could become broken and disordered for certain contact angles. This was confirmed in the experiments by \cite{chang15}, who also showed that the appearance of a dominant mode in regions where two modes may coexist is hysteretic. That is, the observed mode depends upon the direction of the frequency sweep.

\begin{figure}
\begin{center}
\begin{tabular}{cc}
($a$)  & ($b$)  \\
\includegraphics[width=0.4\linewidth]{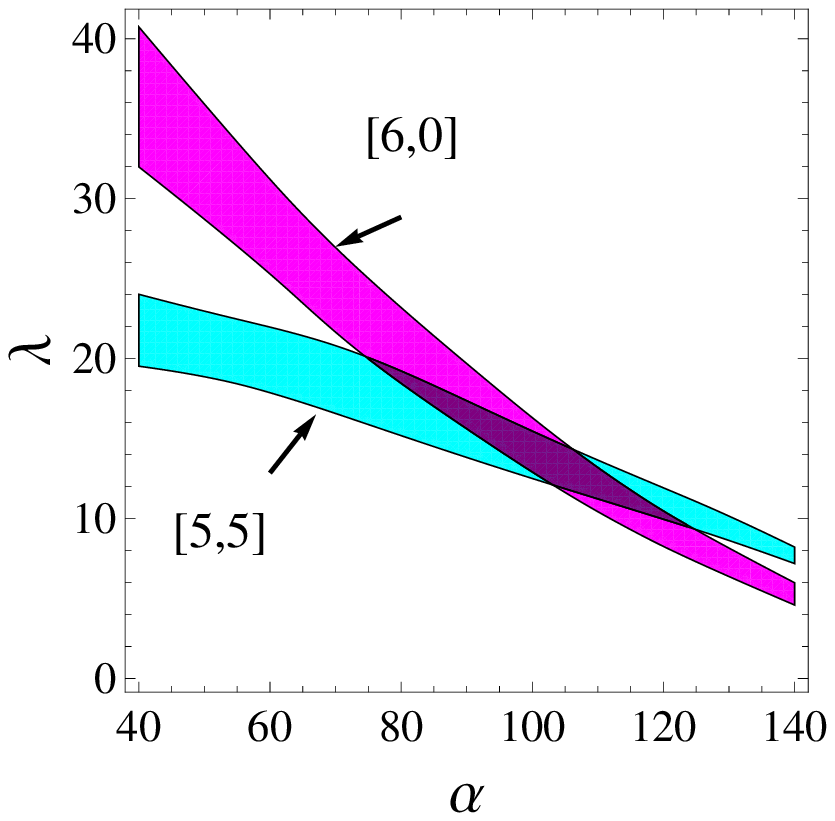}&
\includegraphics[width=0.4\linewidth]{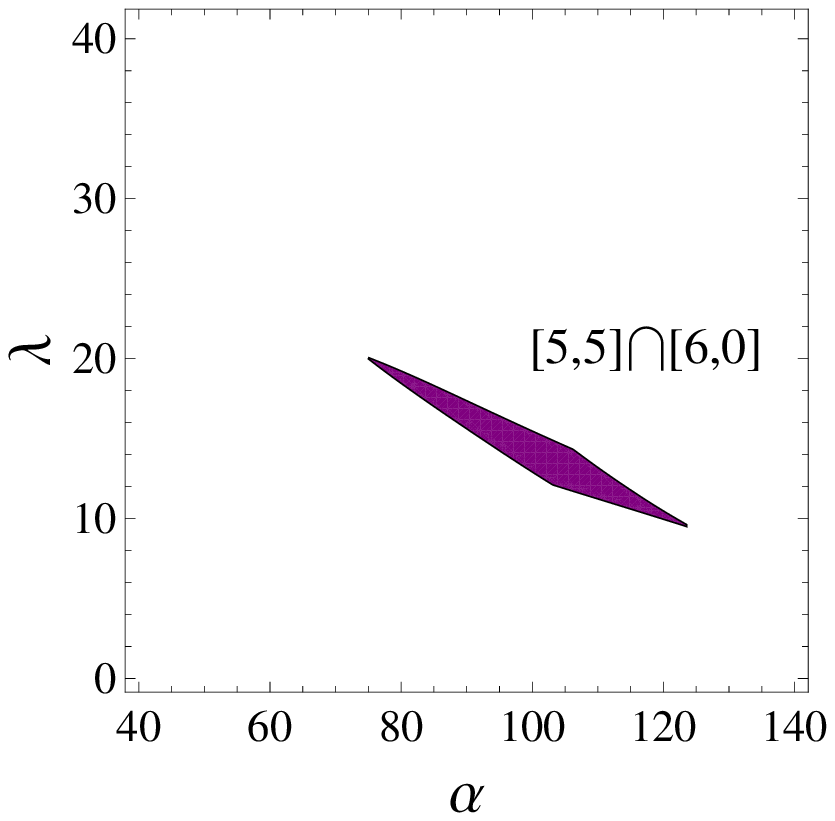}
\end{tabular}
\end{center}
\caption{Modal coexistence: ($a$) frequency $\lambda$ envelopes for the pinned ($\Lambda=\infty$) $[6,0]$ and $[5,5]$ modes against contact-angle $\alpha$ predicts ($b$) domains of coexistence for a drop with $\epsilon=0.0024$.\label{fig:coexist} }
\end{figure}
\begin{figure}
\begin{center}
\begin{tabular}{ccccc}
&$\underline{[2,0]}$ & $\underline{[4,0]}$ & $\underline{[6,0]}$ & $\underline{[8,0]}$  \\
\begin{sideways}\hspace{.6in} $\Lambda=\infty$ \end{sideways} \hspace{-0.15in}&
\includegraphics[width=0.24\linewidth]{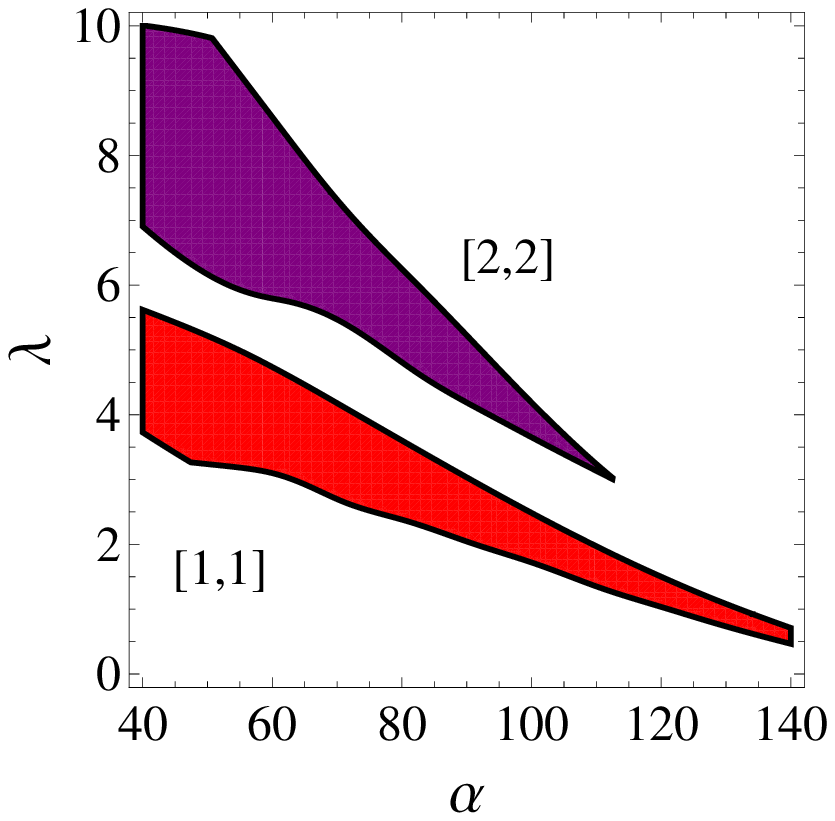}&
\includegraphics[width=0.24\linewidth]{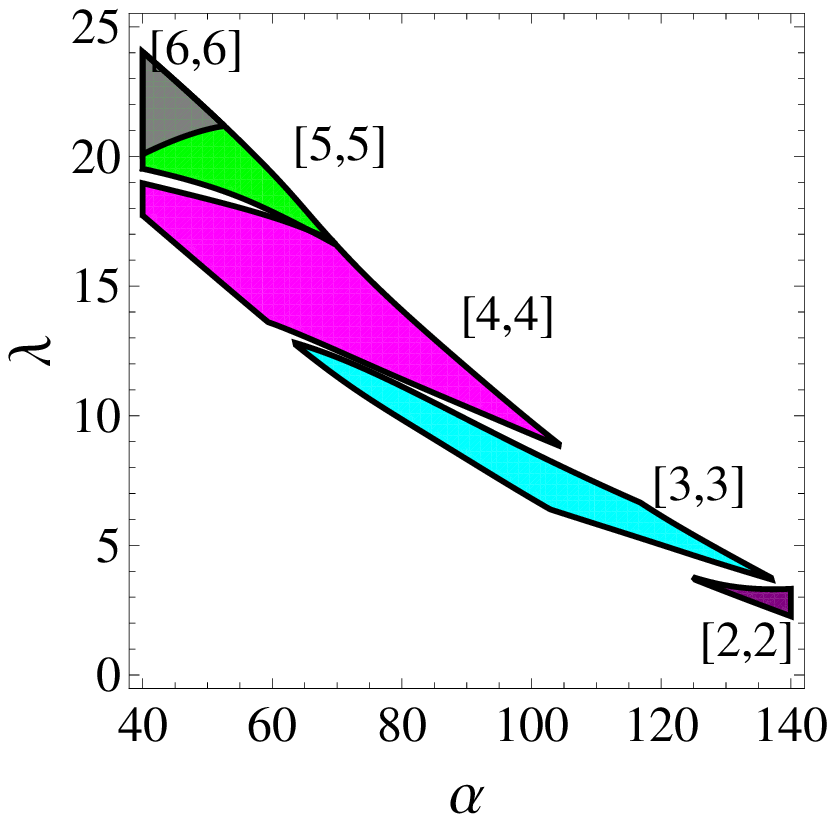} &
\includegraphics[width=0.24\linewidth]{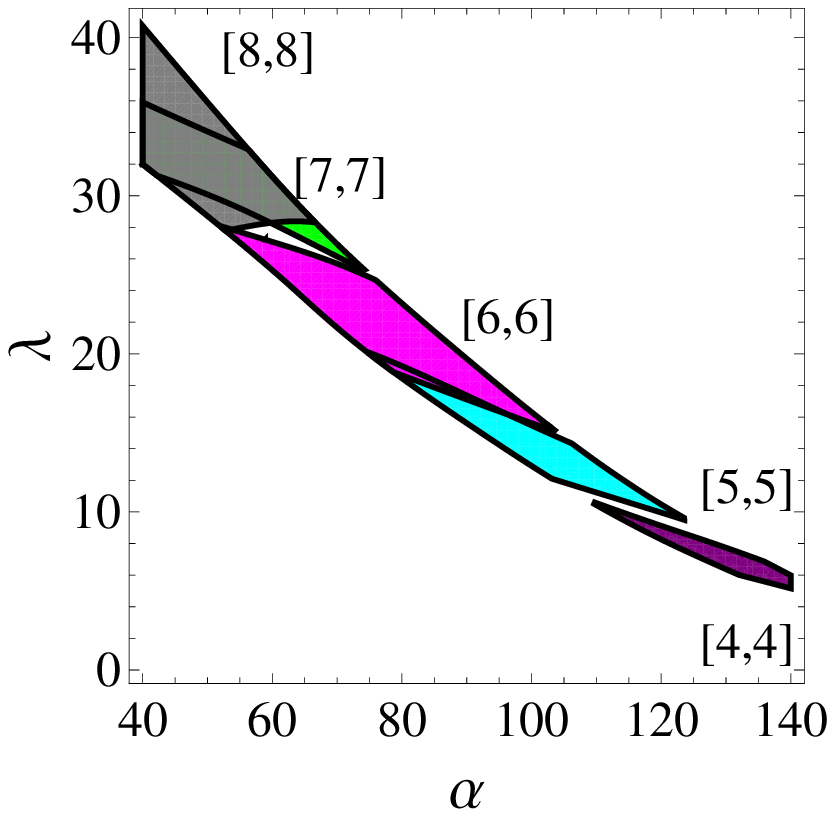} &
\includegraphics[width=0.24\linewidth]{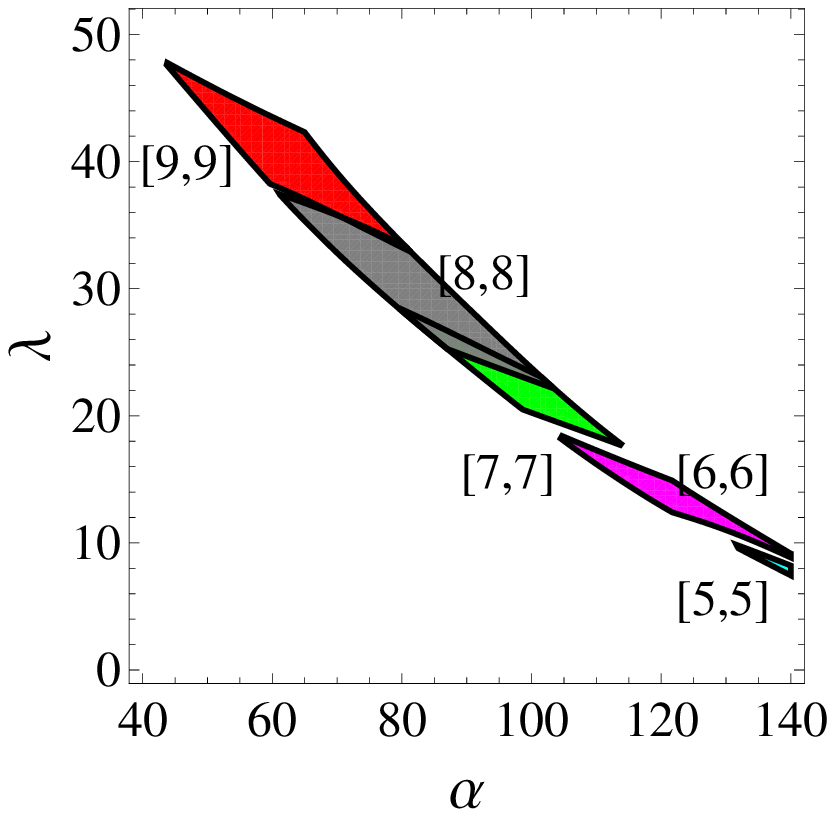} \\
\begin{sideways}\hspace{.5in} $\Lambda=0.1$ \end{sideways} \hspace{-0.15in}&
\includegraphics[width=0.24\linewidth]{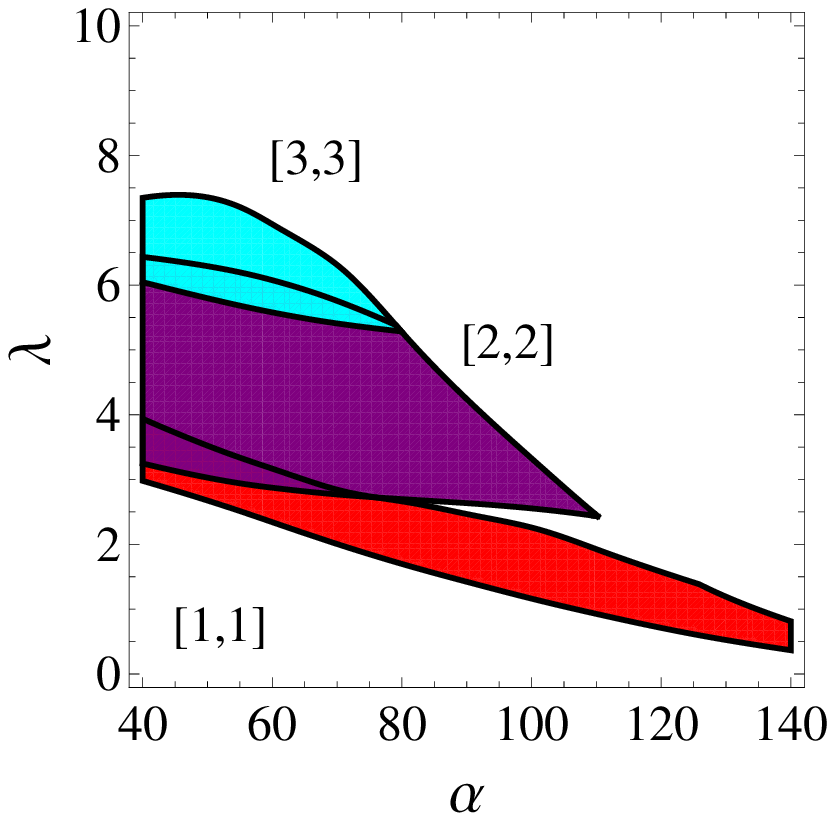}&
\includegraphics[width=0.24\linewidth]{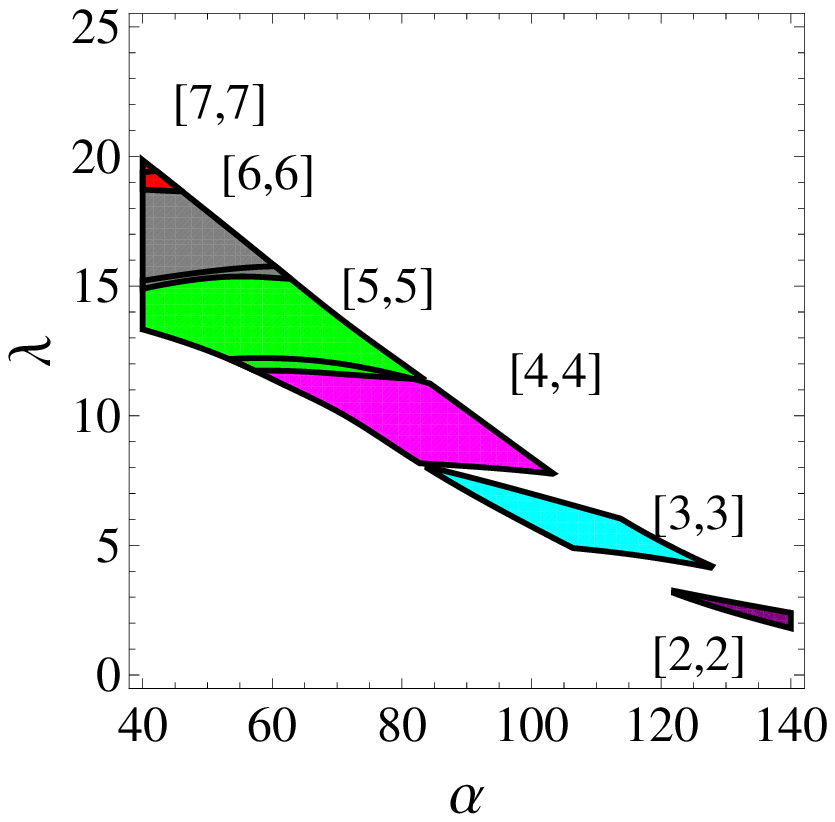} &
\includegraphics[width=0.24\linewidth]{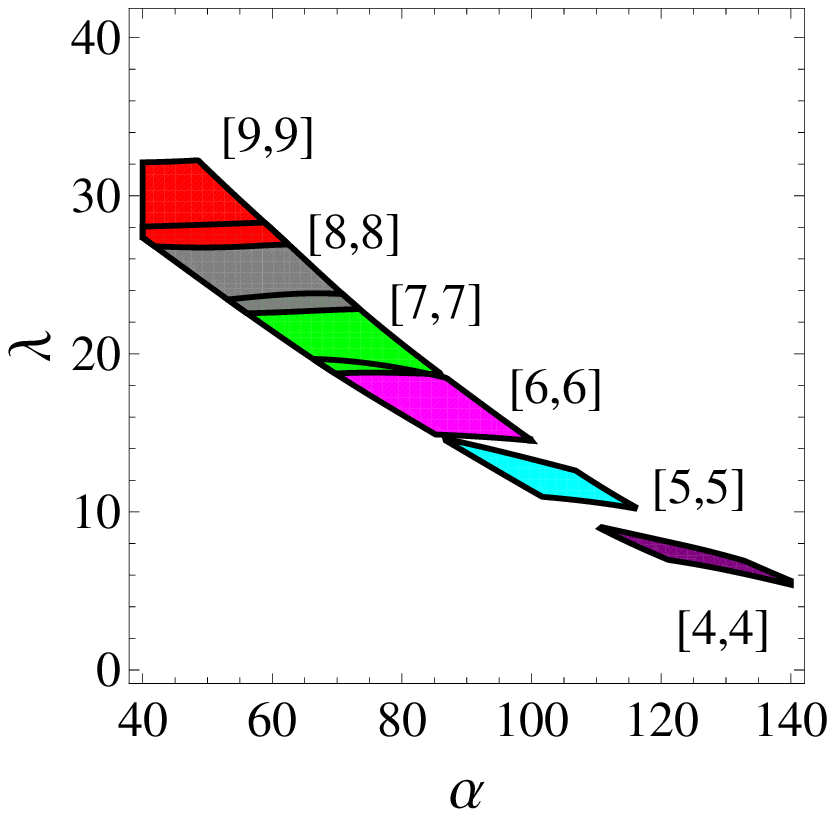} &
\includegraphics[width=0.24\linewidth]{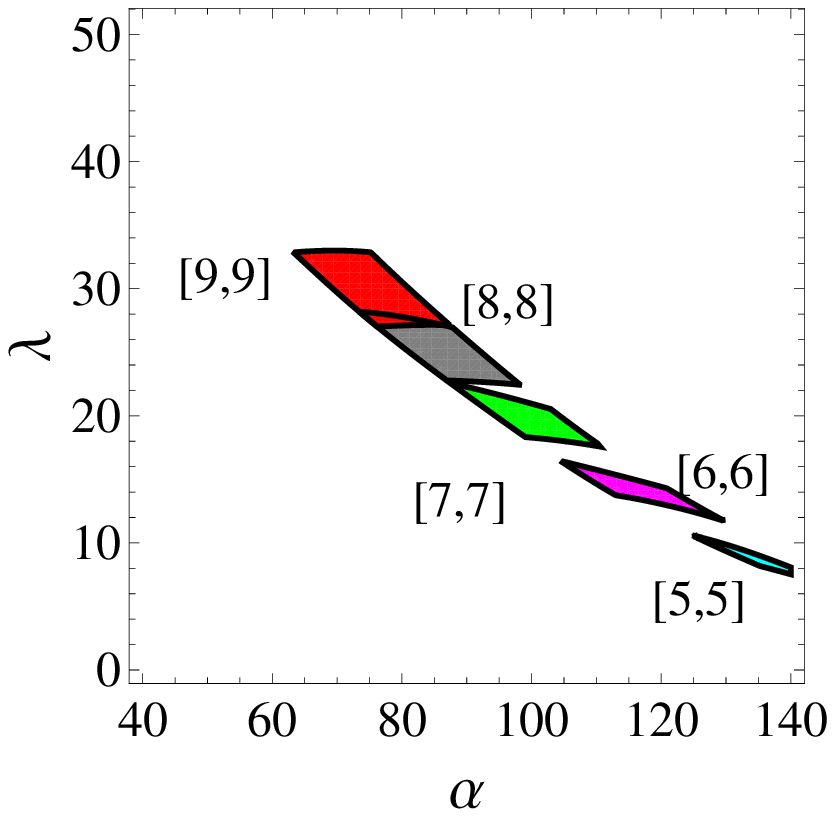} \\
\end{tabular}
\end{center}
\caption{Domains of coexistence for the zonal modes $[k,0]$ mixed with the sectoral modes $[k,k]$ for pinned $\Lambda=\infty$ and finite contact-line mobility $\Lambda=0.1$ disturbances with $\epsilon=0.0024$. Note the different frequency scales between sub-figures. \label{fig:zonsecmix} }
\end{figure}

A primary difference between natural and forced oscillations is that the resonance frequency takes a discrete value for the former and a range of values for the latter. Hence, two modes may coexist over a range of frequencies for the forced problem considered here. Figure~\ref{fig:coexist}($a$) plots the frequency envelopes for the pinned $[6,0]$ zonal and $[5,5]$ sectoral modes against contact angle $\alpha$ for the Ohnesorge number $\epsilon=0.0024$ used in the \cite{chang15} experiments. Modal coexistence is predicted in the region shown in Figure~\ref{fig:coexist}($b$). In general, the domains of coexistence for a pair of modes will depend upon both the contact-line mobility $\Lambda$ and the Ohnesorge number $\epsilon$.  Figure~\ref{fig:zonsecmix} plots the domains of coexistence for a given zonal mode with the sectoral modes, comparing pinned $\Lambda=\infty$ disturbances to those with finite mobility $\Lambda=0.1$. As shown, decreasing $\Lambda$ tends to increase the number of modes that coexist with a given target mode. Figure~\ref{fig:zonsecmix} can be used as a guide to future studies of modal interactions both theoretically and experimentally. For reference, we include additional Figures~\ref{fig:k1sec},\ref{fig:k7sec},\ref{fig:k8sec} that predict domains of coexistence for different target modes in the Appendix.

\section{Concluding remarks}
We have studied the forced oscillations of a partially-wetting sessile drop, whose three-phase contact line obeys a constitutive law relating the contact angle to the contact line speed, sometimes called the Hocking condition. Response diagrams and phase shifts are reported, as they depend upon viscosity $\epsilon$ and contact line mobility $\Lambda$. Modes are distinguished by the wavenumber pair $[k,l]$ and can be excited over a range of frequencies that define a bandwidth. Our predictions compare well against relevant experiments on vibrated sessile drops (c.f. Figures~\ref{fig:FWHM}($b$),\ref{fig:changcomp}), suggesting the predictive nature of our model.

Our focus is on defining regimes or `operating windows' where certain droplet behavior may be observed experimentally or our model developed further. For example, we compute the critical viscosity $\epsilon_c$ (Ohnesorge number) above which it is not possible to observe a specified mode over a range of contact angles, thereby aiding the practitioner in selecting appropriate fluids and droplet volumes (c.f. Figure~\ref{fig:critvisc}). We then show how finite contact line mobility $\Lambda$ leads to Davis dissipation, even in inviscid fluids, and compute the critical mobility $\Lambda_m$ and forcing frequency $\lambda_m$ that generate the largest dissipation (c.f. Figure~\ref{fig:critmob}). Finally, we show that two distinct modes may be simultaneously excited by a single forcing frequency and map these regions of modal coexistence in parameter space for a number of modal pairs in Figure~\ref{fig:zonsecmix}. Modal coexistence may be of importance in mixing applications that rely upon capillary oscillations \citep{mugele2006microfluidic,mampallil11,davoust2013coplanar} and drop atomization \citep{tsai2012ejection,smith} for spray cooling. With regard to modeling, a thorough study of the internal resonances and nonlinear modal interactions \citep{natarajan1987third,henderson1991faraday} in the coexistence domains would help identify the mechanism behind mode selection in related experiments \citep{chang15}.

\bibliographystyle{jfm}
\bibliography{jfmSDp3}

\appendix
\section{Modal coexistence domains}
\begin{figure}
\begin{center}
\begin{tabular}{ccccc}
&$\underline{[3,1]}$ & $\underline{[5,1]}$ & $\underline{[7,1]}$ & $\underline{[9,1]}$  \\
\begin{sideways}\hspace{.6in} $\Lambda=\infty$ \end{sideways} \hspace{-0.15in}&
\includegraphics[width=0.24\linewidth]{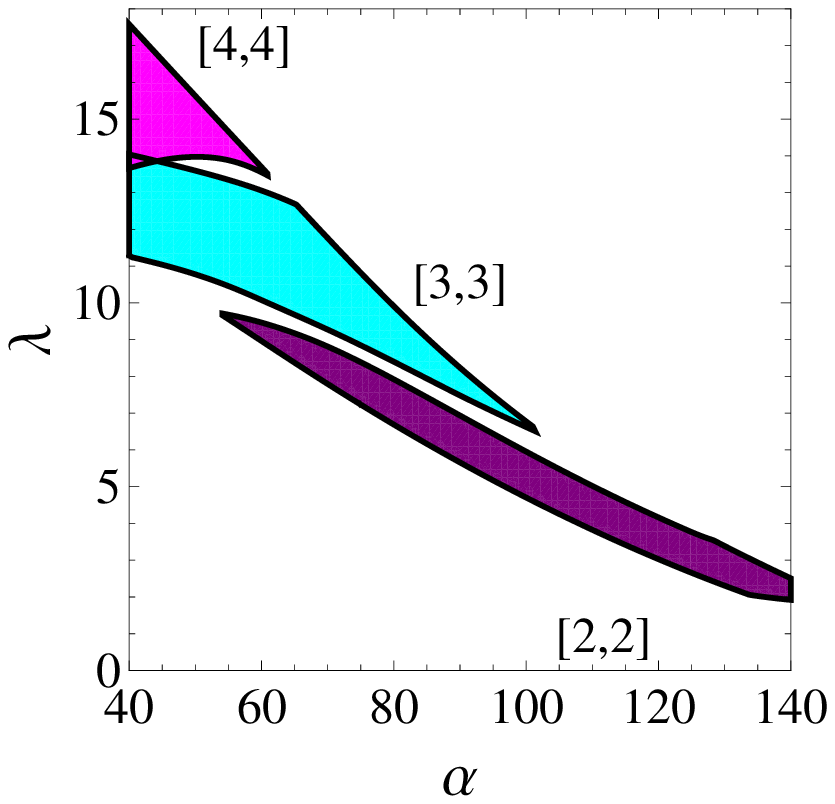}&
\includegraphics[width=0.24\linewidth]{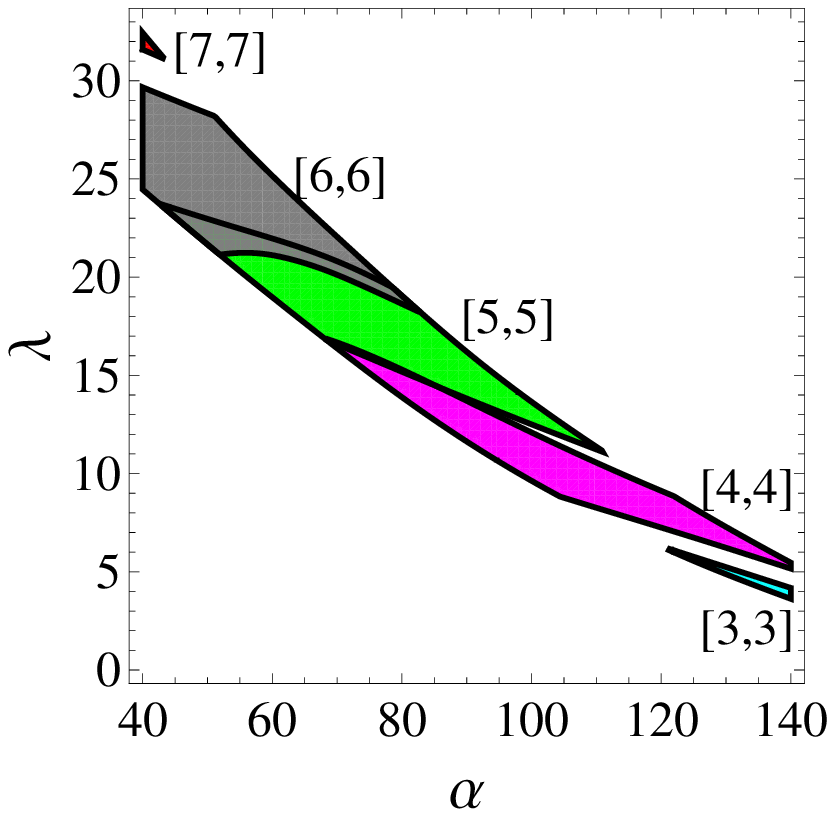} &
\includegraphics[width=0.24\linewidth]{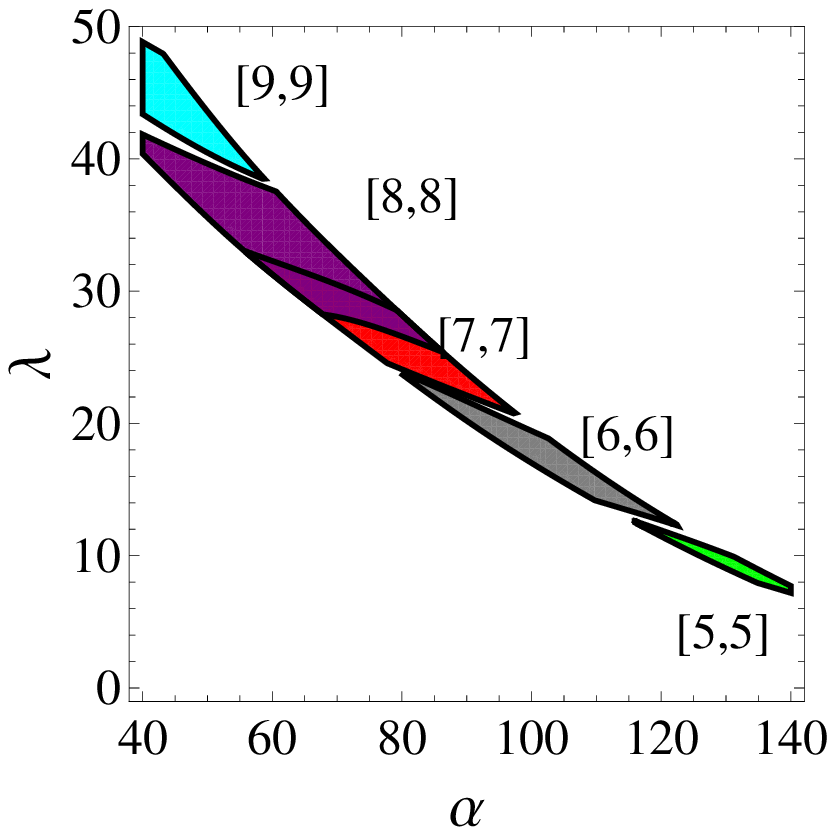} &
\includegraphics[width=0.24\linewidth]{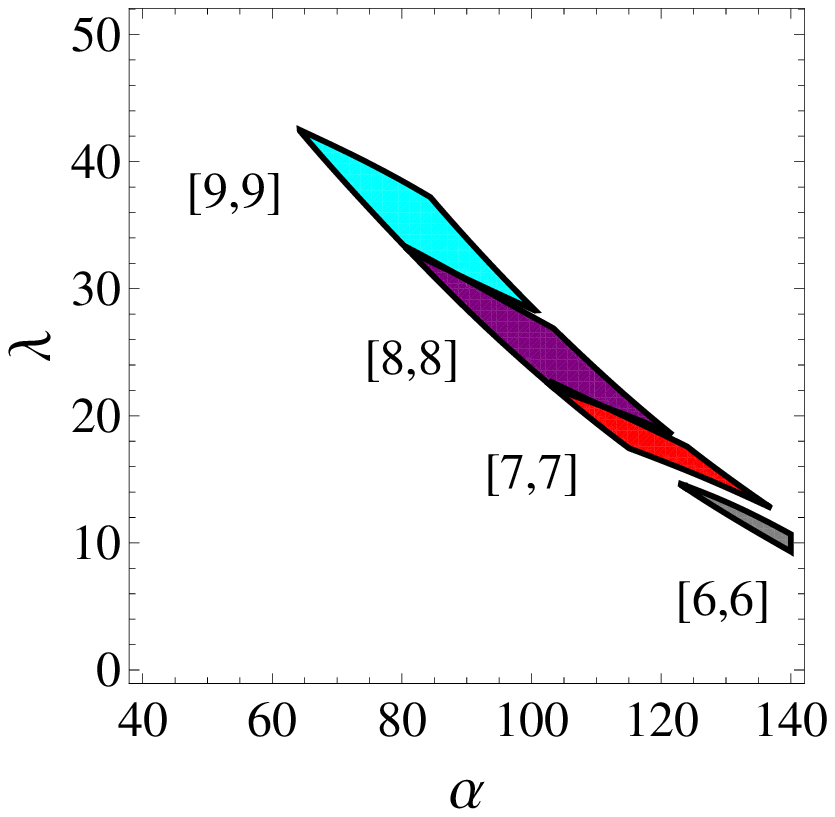} \\
\begin{sideways}\hspace{.5in} $\Lambda=0.1$ \end{sideways} \hspace{-0.15in}&
\includegraphics[width=0.24\linewidth]{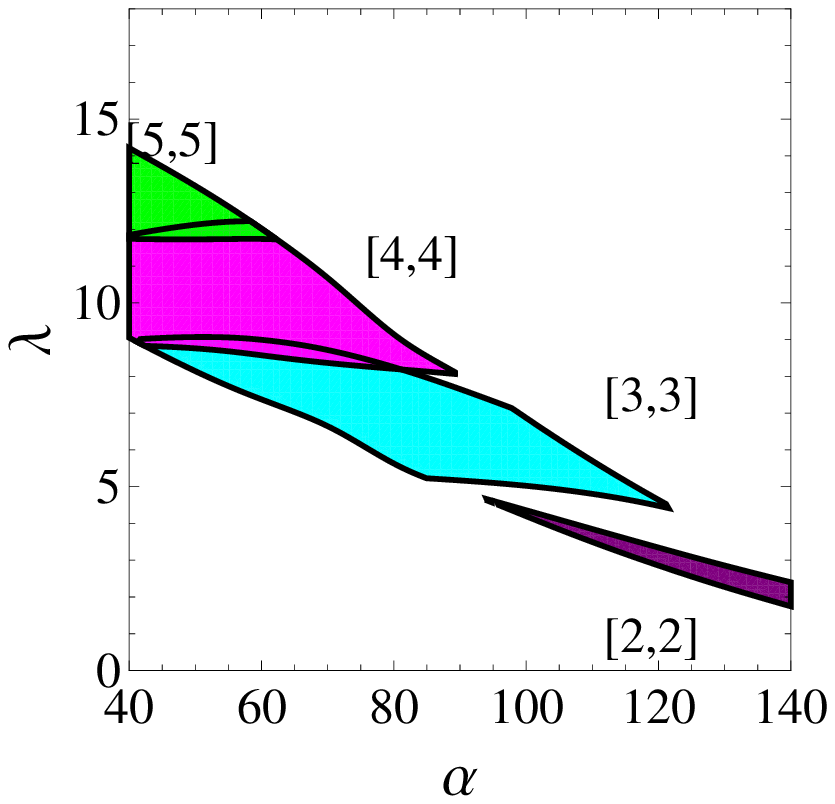}&
\includegraphics[width=0.24\linewidth]{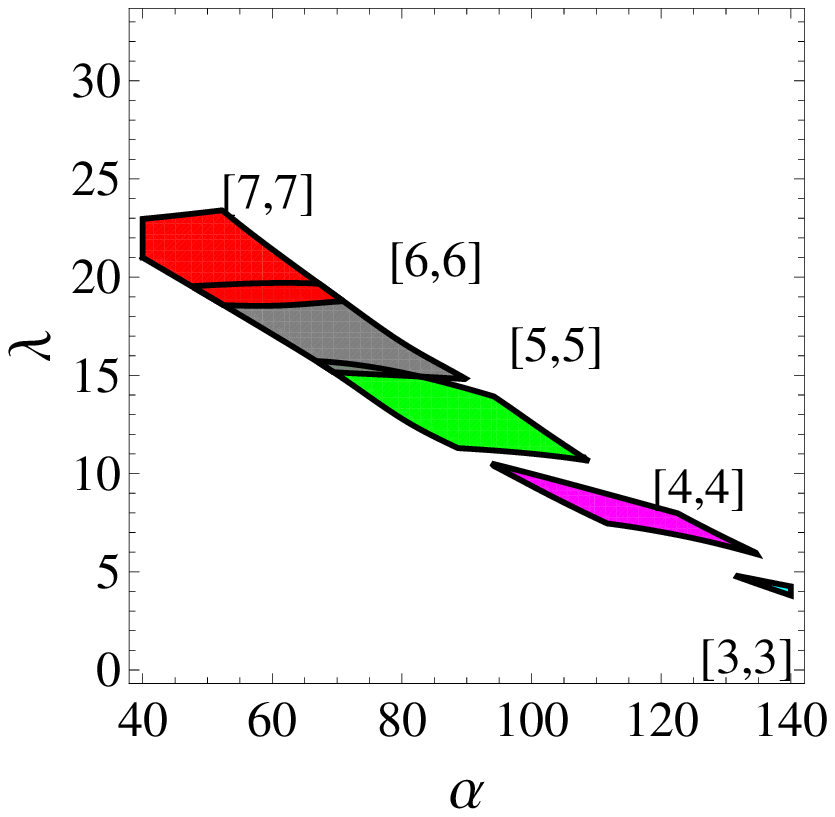} &
\includegraphics[width=0.24\linewidth]{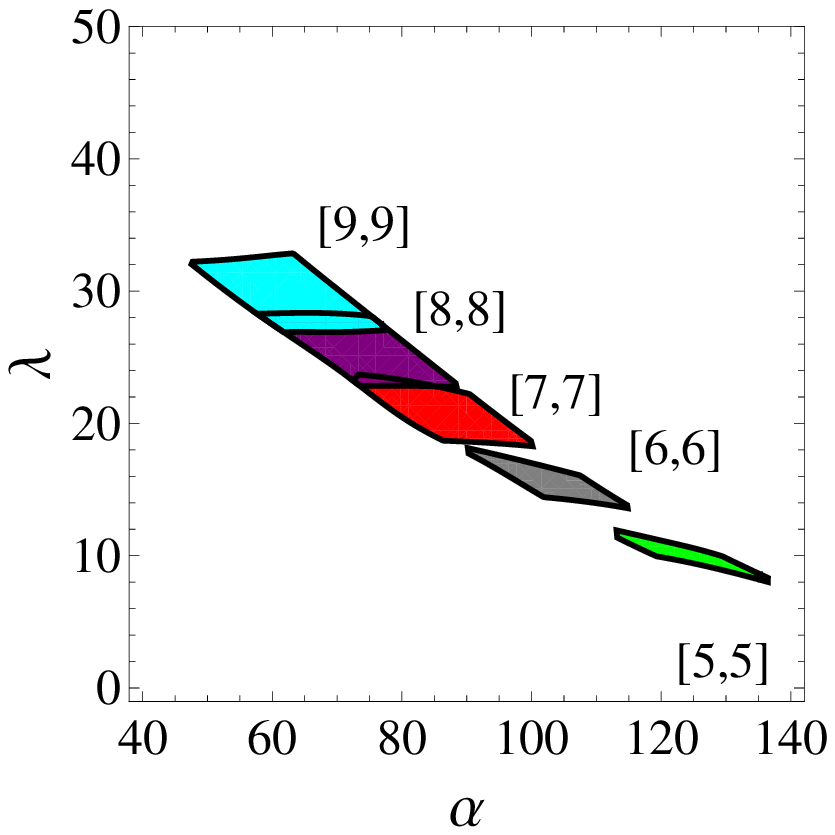} &
\includegraphics[width=0.24\linewidth]{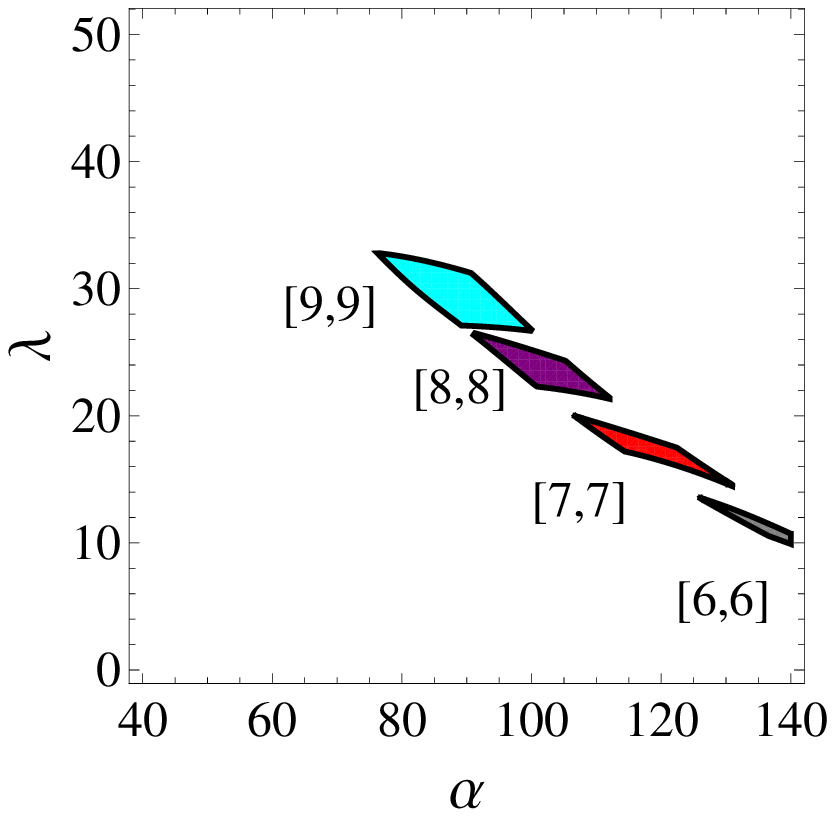} \\
\end{tabular}
\end{center}
\caption{Domains of coexistence for the rocking modes $[k,1]$ mixed with the sectoral modes $[k,k]$ for pinned $\Lambda=\infty$ and finite contact-line mobility $\Lambda=0.1$ disturbances with $\epsilon=0.0024$.  \label{fig:k1sec} }
\end{figure}

\begin{figure}
\begin{center}
\begin{tabular}{cccc}
&$\underline{[7,1]}$ & $\underline{[7,3]}$ & $\underline{[7,5]}$  \\
\begin{sideways}\hspace{.6in} $\Lambda=\infty$ \end{sideways} \hspace{-0.15in}&
\includegraphics[width=0.24\linewidth]{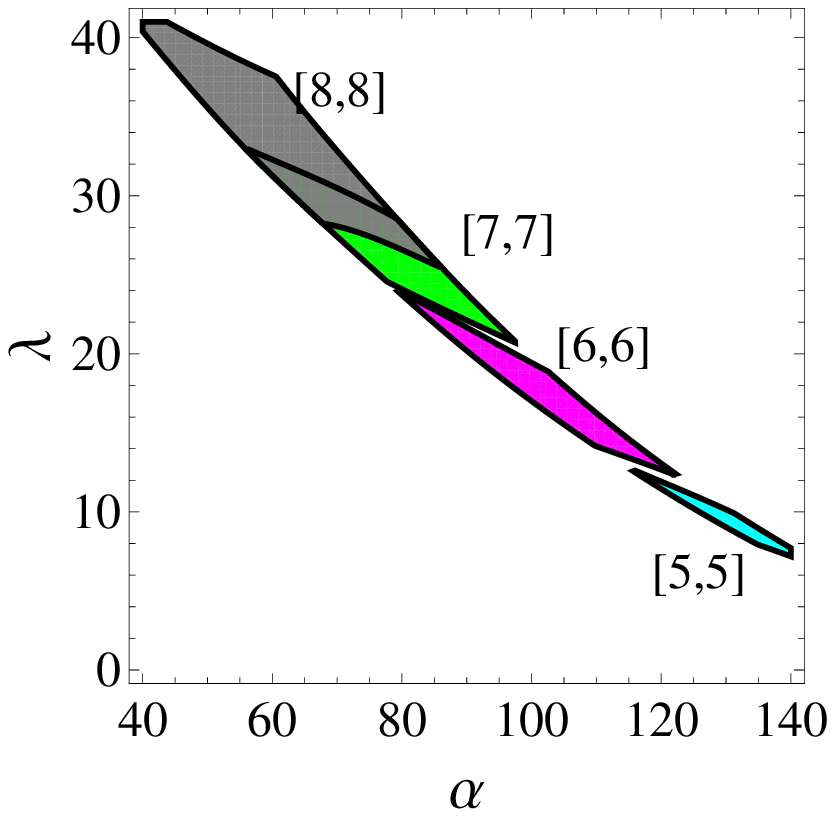} &
\includegraphics[width=0.24\linewidth]{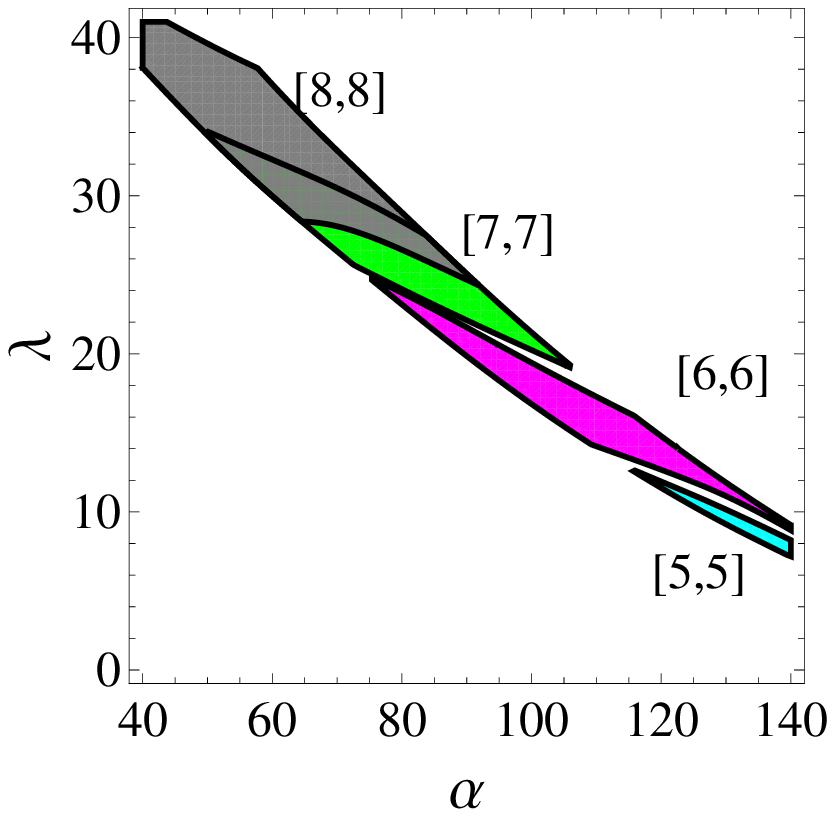} &
\includegraphics[width=0.24\linewidth]{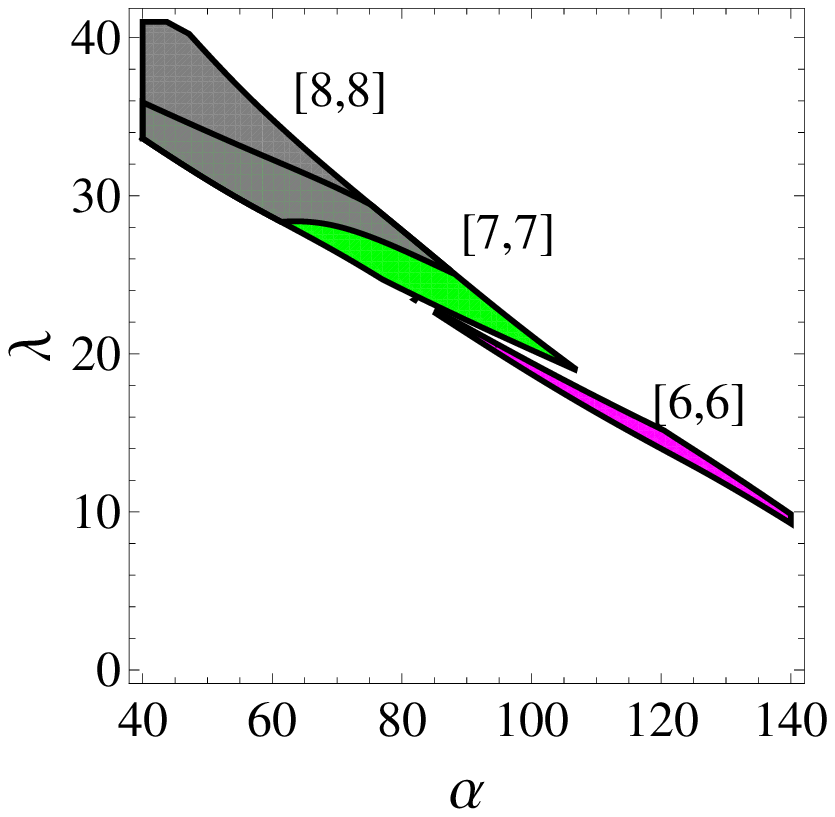} \\
\begin{sideways}\hspace{.5in} $\Lambda=0.1$ \end{sideways} \hspace{-0.15in}&
\includegraphics[width=0.24\linewidth]{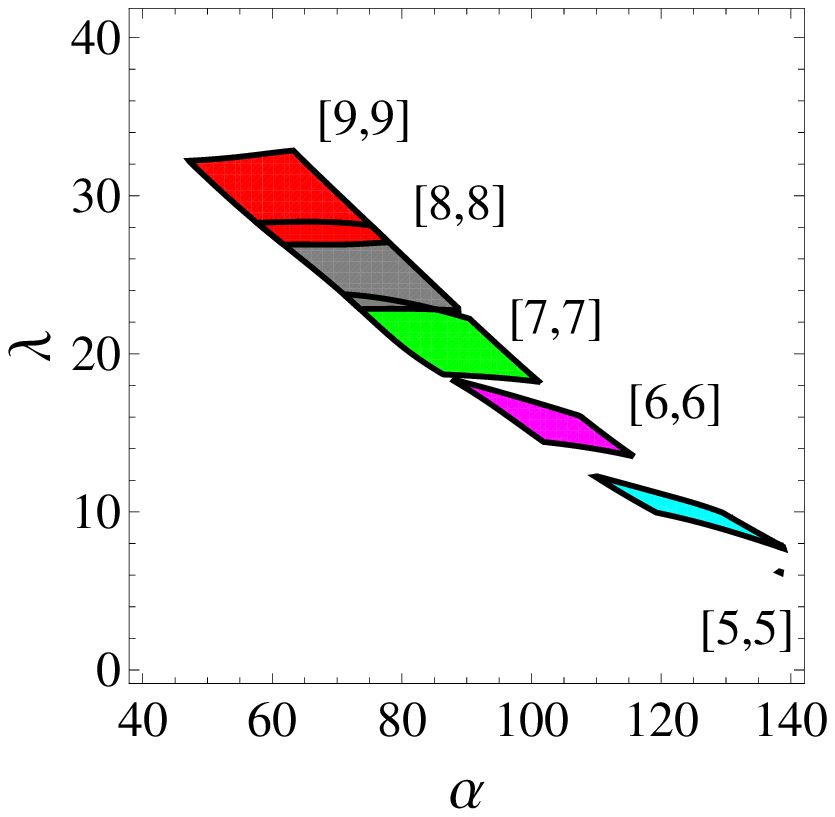} &
\includegraphics[width=0.24\linewidth]{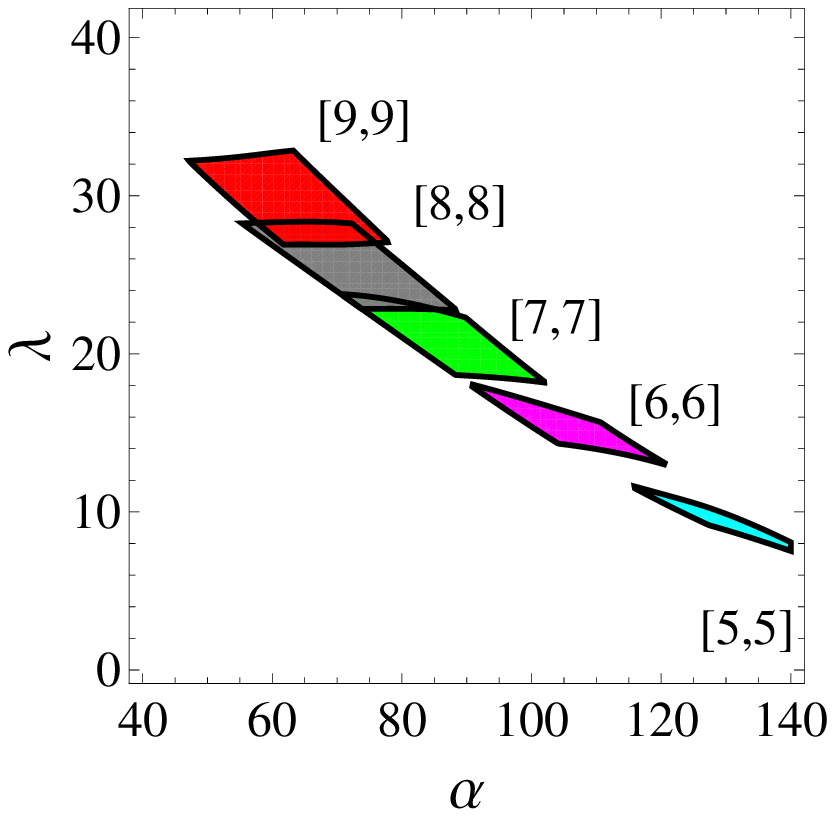} &
\includegraphics[width=0.24\linewidth]{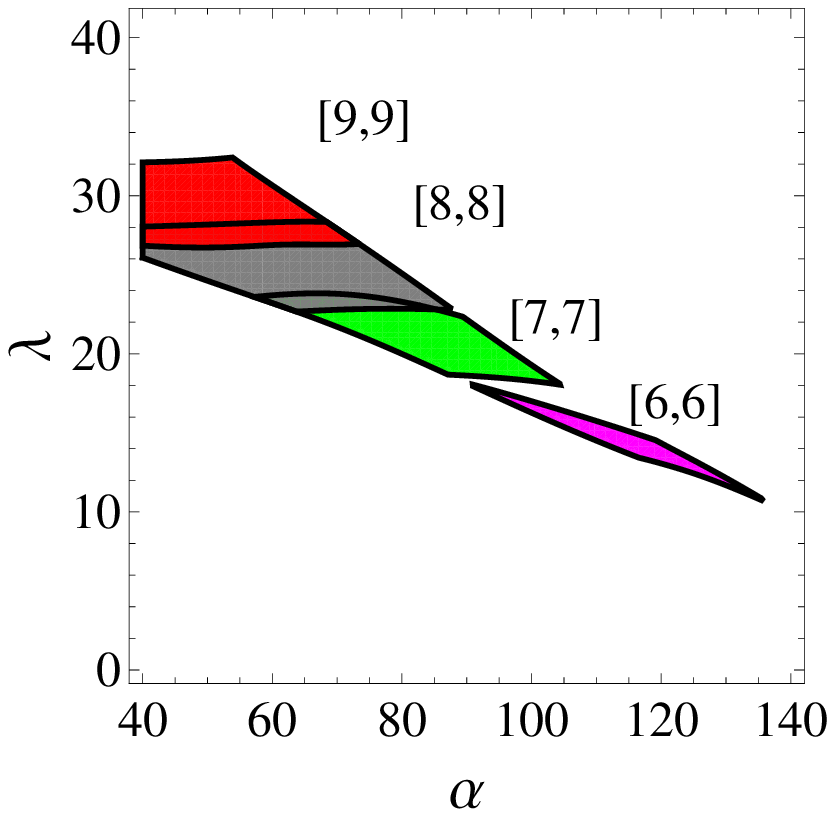} \\
\end{tabular}
\end{center}
\caption{Domains of coexistence for the $[7,l]$ modes mixed with the sectoral modes $[k,k]$ for pinned $\Lambda=\infty$ and finite contact-line mobility $\Lambda=0.1$ disturbances with $\epsilon=0.0024$.  \label{fig:k7sec} }
\end{figure}

\begin{figure}
\begin{center}
\begin{tabular}{cccc}
 & $\underline{[8,2]}$ & $\underline{[8,4]}$ & $\underline{[8,6]}$  \\
\begin{sideways}\hspace{.6in} $\Lambda=\infty$ \end{sideways} \hspace{-0.15in}&
\includegraphics[width=0.24\linewidth]{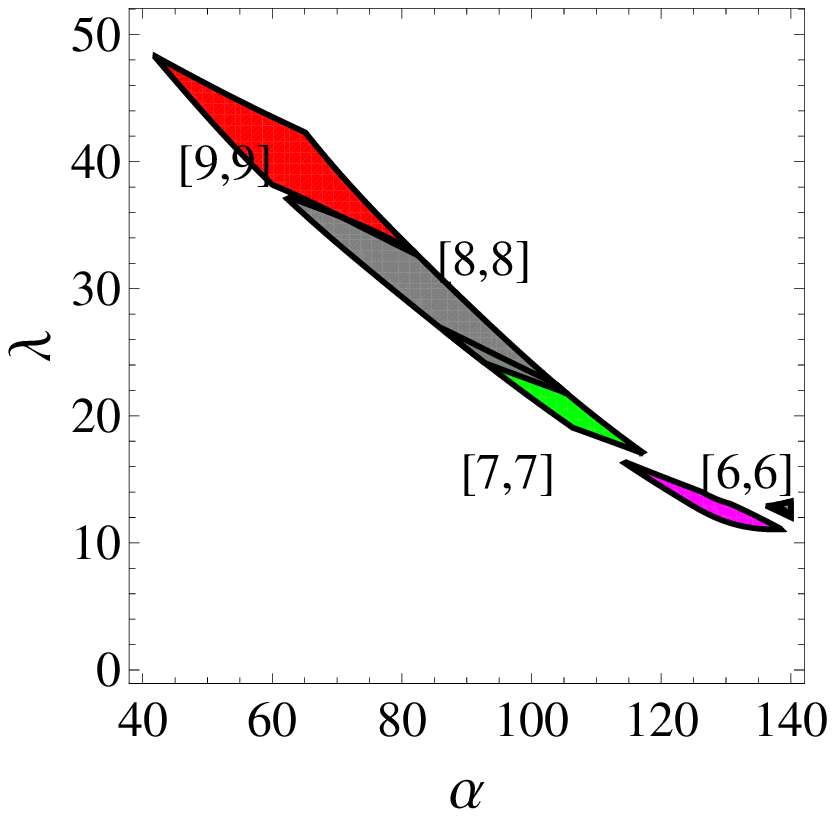} &
\includegraphics[width=0.24\linewidth]{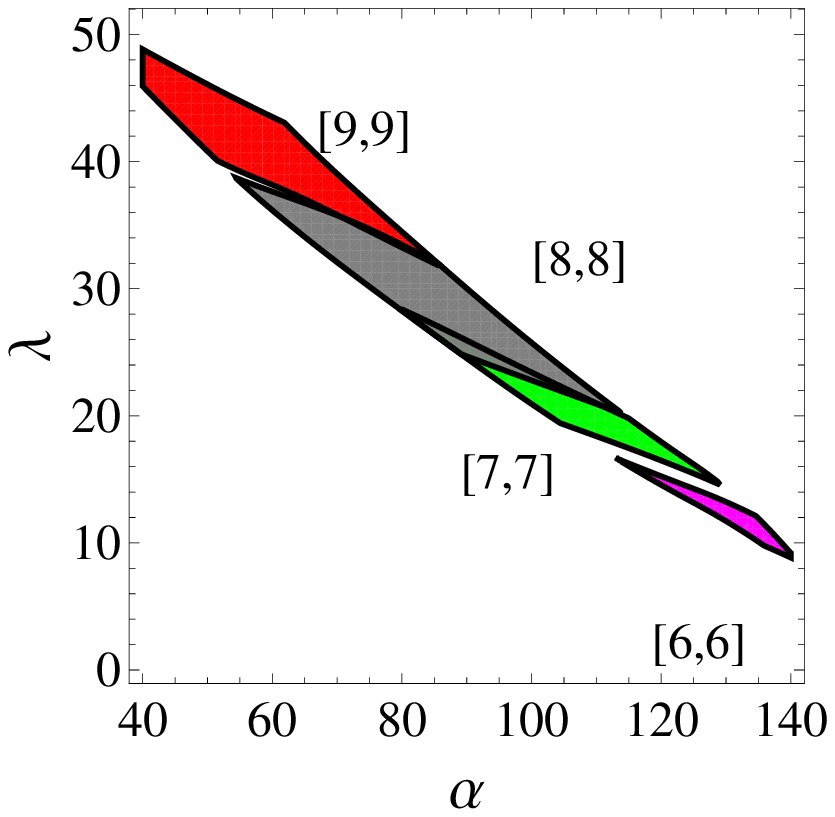} &
\includegraphics[width=0.24\linewidth]{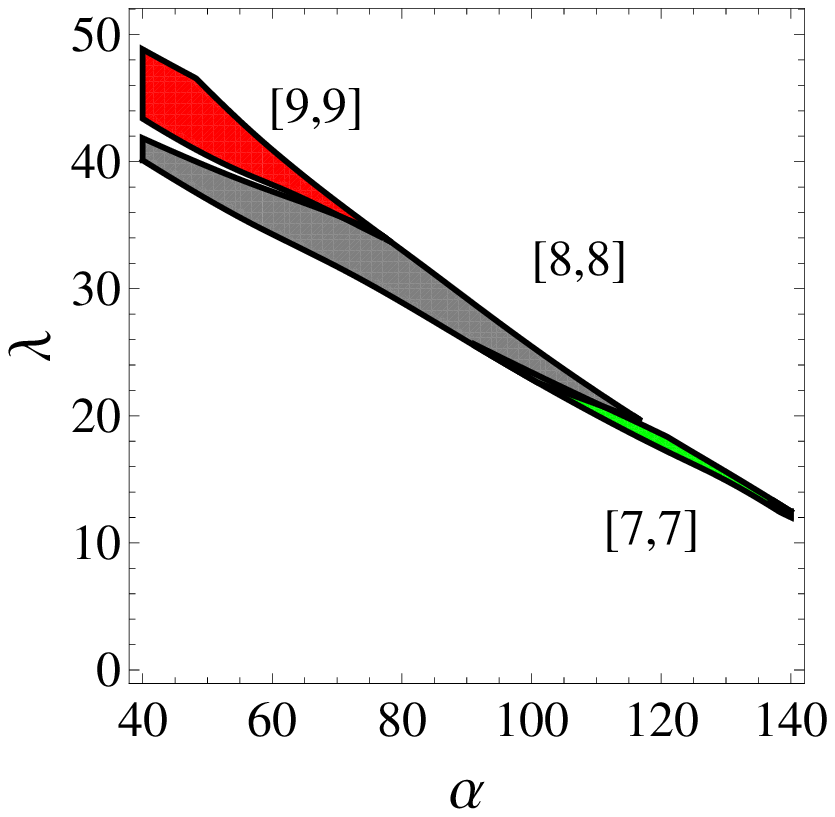} \\
\begin{sideways}\hspace{.5in} $\Lambda=0.1$ \end{sideways} \hspace{-0.15in}&
\includegraphics[width=0.24\linewidth]{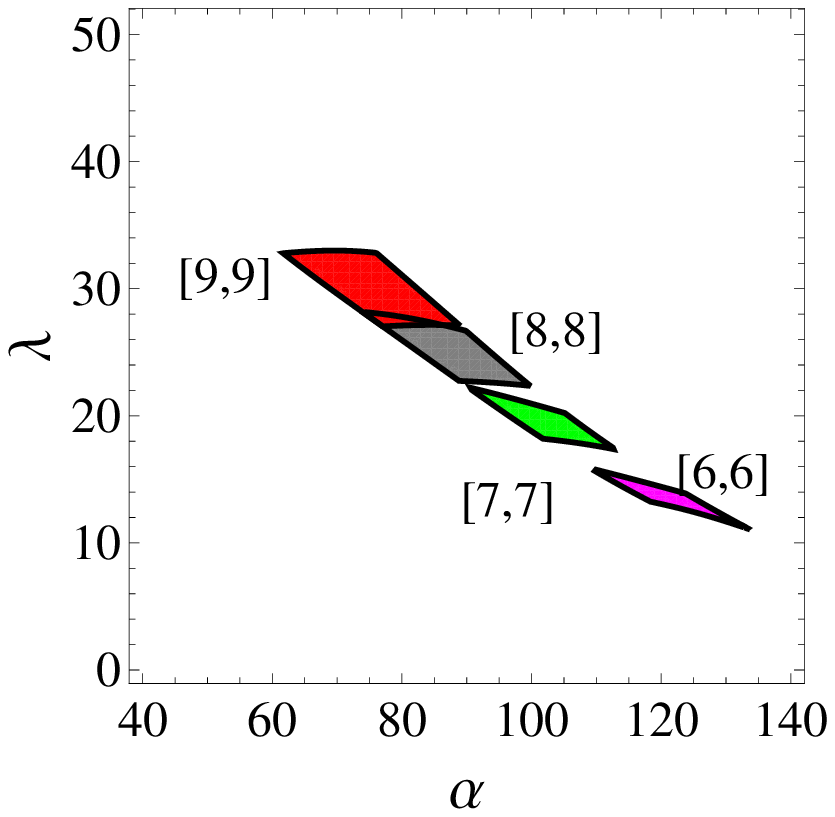} &
\includegraphics[width=0.24\linewidth]{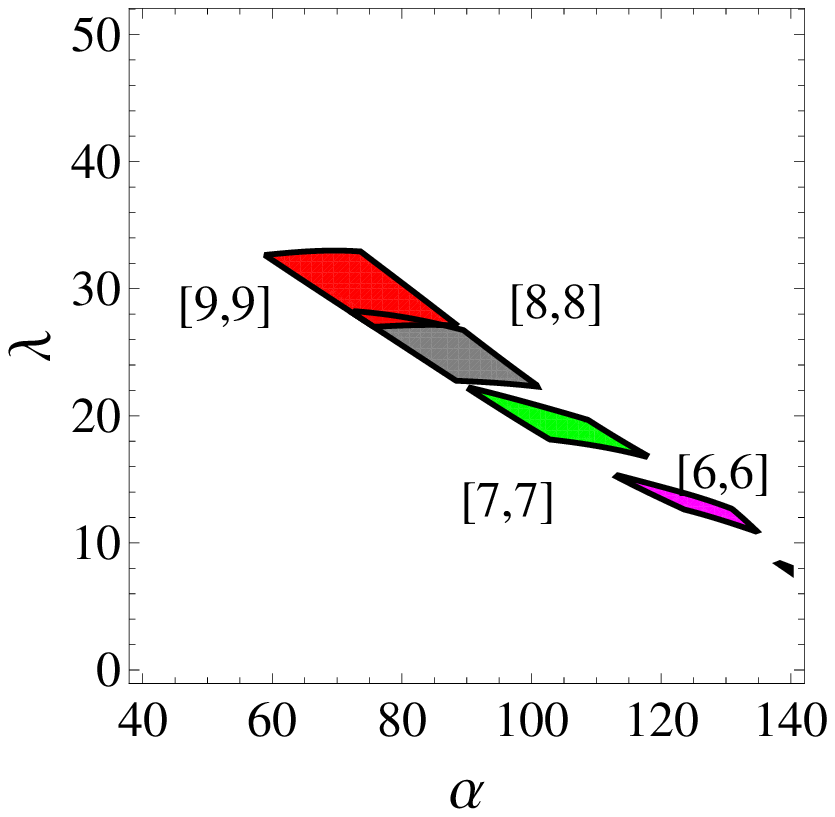} &
\includegraphics[width=0.24\linewidth]{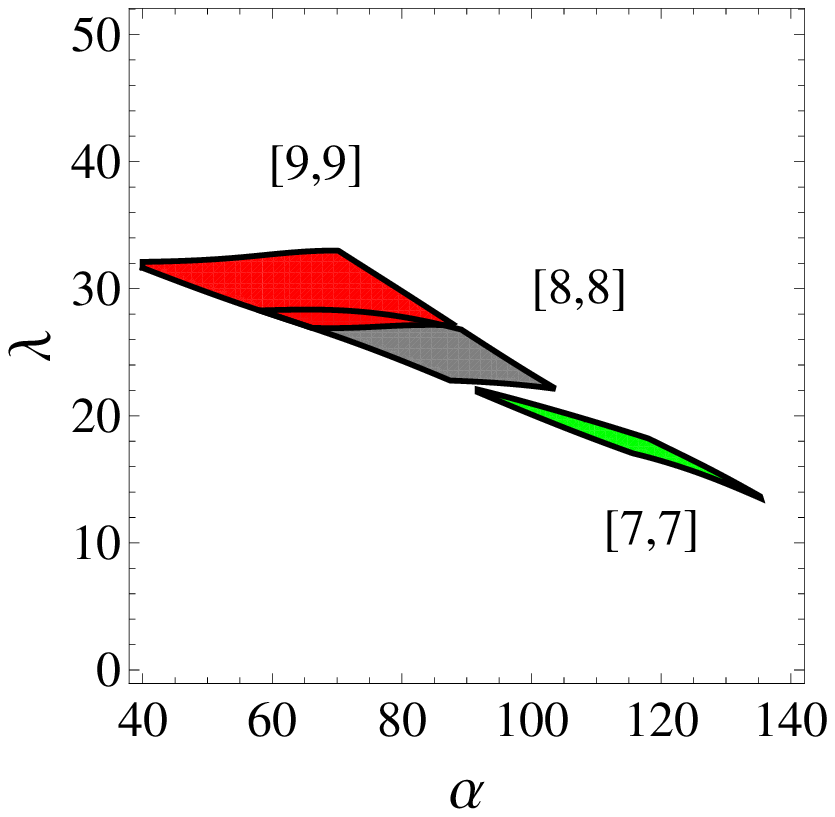} \\
\end{tabular}
\end{center}%
\caption{Domains of coexistence for the $[8,l]$ modes mixed with the sectoral modes $[k,k]$ for pinned $\Lambda=\infty$ and finite contact-line mobility $\Lambda=0.1$ disturbances with $\epsilon=0.0024$. \label{fig:k8sec} }
\end{figure}

Figures~\ref{fig:k1sec}, \ref{fig:k7sec} \& \ref{fig:k8sec} map the regions of modal coexistence for the $l=1,k=7,k=8$ modes with the sectoral modes for pinned $\Lambda=\infty$ and finite mobility $\Lambda=0.1$ disturbances.

%
\end{document}